\documentclass[12pt]{article}
\usepackage[a4paper, left=2cm, right=2cm, top=2cm, bottom=2cm]{geometry}
\usepackage{amsmath,amssymb,color,graphicx, amsfonts}
\usepackage{multirow}
\usepackage{authblk}
\usepackage[utf8]{inputenc}
\usepackage[colorlinks=true
,urlcolor=blue
,anchorcolor=blue
,citecolor=blue
,filecolor=blue
,linkcolor=blue
,menucolor=blue
,pagecolor=blue
,linktocpage=true
,pdfproducer=medialab
]{hyperref}
\numberwithin{equation}{section}

\newcommand{\ETmiss}{{E\!\!\!/}_{\rm T}}

\usepackage{cite}

\begin{document}

\title{Probing EFT breakdown in the tails of $W^+ W^-$ observables}
\date{}
\author[1]{Daniel Gillies}
\author[1]{Andrea Banfi}
\author[2]{Adam Martin}

\affil[1]{Department of Physics and Astronomy, University of Sussex, Brighton BN1 9QH, U.K.}
\affil[2]{Department of Physics, University of Notre Dame, Notre Dame, IN 46556, USA}

\maketitle
\begin{abstract}
  In this letter, we test clipping effective field theory (EFT)
  simulations as a method of ensuring EFT validity. The procedure
  imposes that, at the level of the simulation, the invariant mass of
  a $W^+W^-$ pair $M_{WW}$ is less than the new physics scale
  $\Lambda$. We compare this to two other methods, comparison bin by
  bin of dimension-6 and dimension-8 squared contributions and
  implementing a cut on data. We find that setting $M_{WW} < \Lambda$
  is not strict enough to ensure that the hierarchy of EFT operators
  is respected for dimension-6 and dimension-8 contributions. We also
  show that, even when using a stricter cut on $M_{WW}$, due to
  different correlations between $M_{WW}$ and $M_{e\mu}$ at different
  EFT orders, the bins in $M_{e\mu}$ (the invariant mass of the
  leptons originating from $W$ decays) used in an EFT fit may not
  truly be in the regime of EFT validity when performing a dimension-6
  fit with $M_{WW} < \Lambda$. We also explore the correlations of
  three transverse mass observables: $M_{T1}, M_{T2}$ and $M_{T3}$,
  finding that $M_{T1}$ and $M_{T3}$ follow the $M_{WW}$ distribution
  more closely than $M_{e\mu}$. We present sensitivity studies using
  both the $M_{T3}$ distribution and $M_{e\mu}$ distribution. We test
  implementing an experimental cut on $M_{T3}$ in place of clipping
  the EFT simulation at $M_{WW} < \Lambda$.  We finally comment that
  adding $M_{WW} < \Lambda$ cuts only to the EFT simulation could be
  interpreted as modifying the SMEFT expansion by a form factor and
  could therefore impact the model independence of EFT fits under this
  procedure.
\end{abstract}
\section{Introduction}

Model independent effective field theory (EFT) fits are becoming a
more popular method of interpreting collider
data~\cite{Brivio:2017vri}. The power of these methods is that the
vast space of possible UV complete models is reduced to a
finite-dimensional space of possible directions for new physics at
first order. These corrections are encoded by operators which have
mass dimension $d>4$. Each operator comes with a Wilson coefficient
($c_i$), expected to be $O(1)$, which encodes the importance of that
operator in the new theory. Each operator is suppressed by a factor of
$\Lambda^{d-4}$, where $\Lambda$ is the mass or energy scale of new
physics. The EFT expansion is formally in powers of $E/\Lambda$, where
$E$ is the energy scale of the interaction being considered.  If we
assume that new physics appears at scales much larger than the
energies which can be achieved at current colliders ($E\ll\Lambda$)
then this EFT expansion will be valid and therefore useful predictions
can be extracted.  When $E>\Lambda$ the theory breaks down and is no
longer a useful description. The Standard Model Effective Field Theory
(SMEFT) is one such theory which will be used to constrain collider
data in the near future -- at dimension-6 it contains $61$ operators
(assuming minimal flavour
violation)~\cite{Grzadkowski:2010es}. Performing these fits with total
cross-section data is common, and these measurements can be thought to
exist at some energy scale related to the process being considered
(e.g.\ $M_{Z}$, $M_{H}$, $\Lambda_{\mathrm{EW}}$). However, in order
to better constrain theories, differential cross-section data is also
used. When considering high-energy tails of these distributions, it is
important to place cuts on the energy of the interaction, $E$, such
that the events being used are those which have energy below the EFT
scale being considered. For some observables, the overall energy of an
interaction is not directly resolvable due to missing neutrino
energy. Methods for ensuring EFT validity in differential
distributions are being developed but so far there is no consensus on
a universal best method~\cite{Brivio:2022pyi}.

The production of a pair of EW bosons at the LHC is a very useful
process to probe physics beyond the Standard Model (BSM). In
particular, the contribution of higher-dimensional non-renormalisable
operators tends to grow with energy, so that their effect is expected
to become visible in the tail of suitable kinematic
distributions. Furthermore, these processes give tree-level access to
triple electroweak (EW)-field strength operators (i.e. those
$\propto WWW$)~\cite{ElFaham:2024uop, Azatov:2019xxn, Brivio:2017btx,
  Azatov:2016sqh}.  Importantly, when viewed inclusively, SMEFT $WWW$
does not interfere well with the Standard Model (SM) so the first
place this operator appears is at dimension-6
squared~\cite{Azatov:2019xxn, Degrande:2023iob}. This makes studies
aimed at $WWW$ particularly susceptible to contamination from
dimension-8 and other higher order effects~\cite{ElFaham:2025fow,
  Martin:2023tvi}. One such diboson process is $W^+W^-$ ($WW$)
production which as been measured at the LHC in several
analyses~\cite{ATLAS:2025dhf, CMS:2025nnv, CMS:2020mxy, ATLAS:2019rob,
  CMS:2018zzl, ATLAS:2017uhp, ATLAS:2017bbg, ATLAS:2017zuf,
  ATLAS:2017jag}.

This process is technically more challenging to
describe than others, due to the need to restrict the phase space
available to accompanying jets in order to suppress the otherwise
overwhelming top-antitop background. The quantity that is mostly
sensitive to high energy BSM physics would be the distribution in the diboson
invariant mass $M_{WW}$. However, this quantity is not accessible at
colliders (due to the missing neutrino energy), so appropriate proxies
need to be identified. Typically, one sets constraints on
higher-dimensional operators by looking at (possibly alongside other angular distributions)
 the distribution in $M_{e\mu}$, the invariant mass of the electron and the muon arising
from the decay of the two $W$ bosons. This has been done in several EFT analyses of the 
channel~\cite{ATLAS:2025dhf, CMS:2025nnv, Bellan:2021dcy, Banerjee:2024eyo, Bellm:2016cks, Falkowski:2016cxu}.
However, it has already been shown previously that $M_{e\mu}$ does not correlate well with $M_{WW}$
and so should not be used directly to set EFT validity (e.g. implementing a cut $M_{e\mu} < \Lambda/2$ on data)~\cite{Falkowski:2016cxu}.

In previous work, we noted that, since $WW$ production via gluon
fusion is very suppressed in the SM, the interference between
dimension-8 $GGWW$ EFT operators and the SM can be considered
negligible in the large $M_{e\mu}$ limit~\cite{Gillies:2024mqp}. We
used this fact to argue that EFT fits including the square of
dimension-6 operators can be safely performed (for the $gg$ channel)
without including the interference of the SM with dimension-8
operators, which are formally of the same order ($1/\Lambda^4$).

This argument holds provided the EFT expansion itself is valid.  The EFT
regime is formally established by having $M_{WW} \ll \Lambda$. However,
as mentioned earlier, this quantity is not available at colliders.
We directly compared, bin by bin, the squared contributions from the leading bosonic 
dimension-6 and dimension-8 operators to determine, for a given value of 
$\Lambda$, the range of $M_{e\mu}$ bins over which the EFT expansion remains convergent.
 The procedure
involved including only the bins where the largest dimension-6 squared
contribution ($\sigma^{(6)}$) was at least twice as large as the largest dimension-8
contribution ($\sigma^{(8)}$) or:
\begin{equation}
  \label{eq:empirical_lambdamin_derivation}
  \frac{(2\,\mathrm{TeV})^8\ \sigma^{(8)}_{\Lambda=2\,\mathrm{TeV}}}{\Lambda^8} = \frac{1}{2}\frac{(2\,\mathrm{TeV})^4\ \sigma^{(6)}_{\Lambda=2\,\mathrm{TeV}}}{\Lambda^4}\,,
\end{equation}
in the assumption that all the coefficients of the higher-dimensional
operators were of order one~\cite{Gillies:2024mqp}. To obtain
numerical estimates, we computed cross-sections at a reference value
of $\Lambda=2\,$TeV. This led to the condition that a bin should only
be used when the EFT scale $\Lambda$ being probed is greater than
$\Lambda_{\min}$, where
\begin{equation}
  \label{eq:empirical_lambdamin}
  \Lambda_{\min} = (2\,\mathrm{TeV})\left(2\times\frac{\sigma^{(8)}_{\Lambda=2\,\mathrm{TeV}}}{\sigma^{(6)}_{\Lambda=2\,\mathrm{TeV}}}\right)^\frac{1}{4}.
\end{equation}
An alternative strategy employed elsewhere does not consider the
$M_{e\mu}$ bins at all. Rather, the dimension-6 interference and
squared contributions are computed with a leading-order (LO) event
generator with the constraint that
$M_{WW} <
\Lambda$~\cite{ATLAS:2025dhf,Brivio:2022pyi,Falkowski:2016cxu}. Adding
this constraint is formally equivalent to modifying the SMEFT Feynman
rules for a dimension-6 operator with Wilson coefficient $c_i$ as
follows:
\begin{equation}
  \label{eq:eft_with_cut}
  \frac{c_i}{\Lambda^2} \rightarrow \frac{c^\prime_i}{\Lambda^2}\,\Theta(1 - p^2/\Lambda^2)\,,
\end{equation}
The step function is applied only to the simulation of the dimension-6
contribution, not the SM contribution or the data, which means that the we are no
longer strictly fitting SMEFT dimension-6 operators. In fact, if we
replace the step function in eq.~\eqref{eq:eft_with_cut} with a smooth
form factor $\mathcal{F}(p^2/\Lambda^2)$ that suppresses higher values
of $p^2$, this would correspond to an infinite series of higher
dimensional operators, which means we are no longer truncating the EFT
expansion at dimension-6.  The step function in
eq.~\eqref{eq:eft_with_cut} is not smooth and does not permit an
expansion in $p^2/\Lambda^2$.  However, it is hoped that fitting
$c^\prime_i$ to the data should give some
indication of what value $c_i$ could be constrained to in the EFT
valid regime. A cleaner method might be to still fit $c_i$ to the data
with the standard dimension-6 SMEFT operator but to only use regions
of phase space (or bins) where including the step function has a small impact.
In this paper, we do not consider this latter method
further and we just compare to what is currently used in recent experimental
analyses, which is to fit on $c^\prime_i$~\cite{ATLAS:2025dhf}. 

There, it is assumed that, with the cut on $M_{WW}$ at the generator
level, the dimension-6 operators will only produce meaningful
contributions in those $M_{e\mu}$ bins for which $M_{WW}$ is within
the EFT validity regime. However, with either of the two above
methods, there is no guarantee that the selected $M_{e\mu}$ bins will
not receive contributions from higher dimensional ($8$ or higher)
operators at larger values of $M_{WW}$. This is because these
higher-dimensional operators can grow faster with energy than
dimension-6 operators, so events with larger values of $M_{WW}$ can
have larger cross-sections, and hence have a bigger impact at lower
values of $M_{e\mu}$.

More precisely, we wish to understand what the expected value of
$M_{WW}$ is for a given $M_{e\mu}$ -- i.e
$\mathbb{E}\left[M_{WW} \mid M_{e\mu} \right]$.  This quantity gives
us an indication of the minimum EFT scale $\Lambda$ that is required
for each bin $M_{e\mu}$. If this quantity changes as we include higher
orders in the EFT, then the $M_{e\mu}$ bins which are included in the
dimension-6 squared prediction could have large higher-order EFT
contributions which are not accounted for.

In this letter, we want to compare the method of fitting $\frac{c^\prime_i}{\Lambda^2}\,\Theta(1 - p^2/\Lambda^2)$
to alternatives and interpret the outcome by explicitly
studying the correlation between $M_{e\mu}$ and $M_{WW}$. For this study,
we choose the bosonic dimension-6 and dimension-8 operators that give
the largest contributions to the $gg$-channel. They are
\begin{equation}
  \label{eq:gg-CP-even}
  \begin{split}
    \mathcal{O}^{(6)}_{GH} & \equiv H^\dagger H\,G_{\mu\nu}^a G^{a,\mu\nu}\,, \\
    \\
\mathcal{O}^{(8)}_3 & \equiv G_{\mu\nu}^a \tilde G^{a,\mu\nu} W^{I,\rho\sigma} \tilde W^{I}_{\rho\sigma}\,.
  \end{split}
\end{equation}
where $G_{\mu\nu}^a$ and $W^{I}_{\rho\sigma}$ are the gluon and $W$
field-strength tensors, and
$\tilde T_{\mu\nu}=\frac 12 \epsilon_{\mu\nu\rho\sigma}T^{\rho\sigma}$ the
dual of any tensor $T^{\mu\nu}$.  Finally, we will also study the
correlation of $M_{WW}$ to other transverse-mass observables which
should more closely follow $M_{WW}$, to see if observables which are
more correlated to $M_{WW}$ give better constraints than those
obtained with $M_{e\mu}$.

$WW$ production in the SM is dominated by $q\bar q$-initiated amplitudes,
whilst this study focuses exclusively on bosonic operators which enter
into the $gg$-channel. The philosophy advocated for in this paper is
that, for the EFT to be valid, one should check its validity for all
processes, not just those which dominate the cross-section. Not doing so assumes that
the expansion for $q\bar q$ operators is different to that of the $gg$
operators (i.e.\ that $gg$ operators are suppressed relative to
$q\bar q$ operators as $\Lambda_{gg} \gg \Lambda_{q\bar q}$ or
$(c_i)_{q\bar q} \gg (c_i)_{gg}$).  Note that we ourselves do not
consider here the fermionic and $q\bar q$-operators as well as some
higher order SM effects, therefore the constraints from this paper
should be treated as purely illustrative.
		
\section{Comparison of EFT Validity Methods}
In this section, we want to compare methods of ensuring fits of Wilson coefficients are performed
only on bins which are within the EFT validity regime for a given $\Lambda$. At the very least, this should require that
dimension-8 squared contributions are negligible compared to the dimension-6 squared contributions
for the kinematical values of the data points.

We consider a simplified model with just the most significant dimension-6 and dimension-8 operators as
 in eq.~\eqref{eq:gg-CP-even} as BSM contributions:
\begin{equation}
  \label{eq:effective-Lkappa}
  \mathcal{L}\supset \mathcal{L}_{\mathrm{BSM}} =\frac{c^{(6)}_{GH}}{\Lambda^2}G_{\mu\nu}^a G^{a,\mu\nu}+\frac{\tilde c^{(6)}_{GH}}{\Lambda^2}G_{\mu\nu}^a \tilde G^{a,\mu\nu}+\sum_i \frac{c^{(8)}_3}{\Lambda^4} G_{\mu\nu}^a \tilde G^{a,\mu\nu} W^{I,\rho\sigma} \tilde W^{I}_{\rho\sigma}\,.
\end{equation}
We can then determine a BSM contribution to the amplitude needed to
  compute physical distributions as:
\begin{equation}
  \label{eq:Mgg}
  \mathcal{M}_{\mathrm{BSM}}  = \frac{c^{(6)}_{GH}}{\Lambda^2}  \mathcal{M}^{(6,\,gg)}_g + \frac{\tilde c^{(6)}_{GH}}{\Lambda^2}  \tilde{\mathcal{M}}^{(6,\,gg)}_g + \frac{c^{(8)}_3}{\Lambda^4} \mathcal{M}^{(8,\,gg)}_3\,.
\end{equation}
These terms may either be squared or interfered with SM $gg$
production.  Both of these contributions at dimension-6 and
dimension-8 are available at LO and next-to-leading logarithmic (NLL)
accuracy (when resummed with a jet-veto $p_{T,\mathrm{veto}}$) in
MCFM-RE~\cite{Gillies:2024mqp, Arpino:2019fmo, Campbell:1999ah,
  Campbell:2011bn, Campbell:2015qma}. For the numerical results in
this letter, we consider $WW$ production at the HL-LHC, with fiducial
cuts based on the ATLAS study in~\cite{ATLAS:2019rob}. These are
summarised in table~\ref{tab:fiducial-cuts}. Unless specified
otherwise, we use the
``NNPDF31$\_$nnlo$\_$as$\_$0118$\_$luxqed$\_$nf$\_$4'' PDF
set~\cite{Bertone:2017bme}. Where mentioned, we use the
``NNPDF40$\_$nnlo$\_$as$\_$01180$\_$nf$\_$4'' PDF set due to better
accuracy in the high-$x$ region~\cite{NNPDF:2021njg}.
\begin{table}[!htbp]
\begin{center}
\begin{tabular}{c|c}
  \hline\hline
  Fiducial selection requirement  & Cut value \\
  \hline\hline
  $p_T^{\ell}$ & $>27\,\mathrm{GeV}$ \\
  $|y_{\ell}|$ & $<2.5$ \\
  $M_{e\mu}$ & $>55\,\mathrm{GeV}$ \\
  $|\vec{p}_T^{\ e}+\vec{p}_T^{\ \mu}|$ & $>30\,\mathrm{GeV}$ \\
  Number of jets with $p_T> 35\,\mathrm{GeV}$ & 0 \\
  $\ETmiss$ & $>20\,\mathrm{GeV}$ \\
  \hline\hline
\end{tabular}
 \end{center}
 \caption{
   Definition of the $WW\rightarrow e\mu$ fiducial phase space, where
   $\vec{p}_T^{\ \ell},y_\ell$ are the transverse momentum and rapidity of either an
   electron or a muon, $M_{e\mu}$ is the invariant mass of the electron-muon
   pair, and $\ETmiss$ is the missing transverse energy.
 }
   \label{tab:fiducial-cuts}
 \end{table}

\begin{figure}[!htbp]
  \includegraphics[width=.5\textwidth]{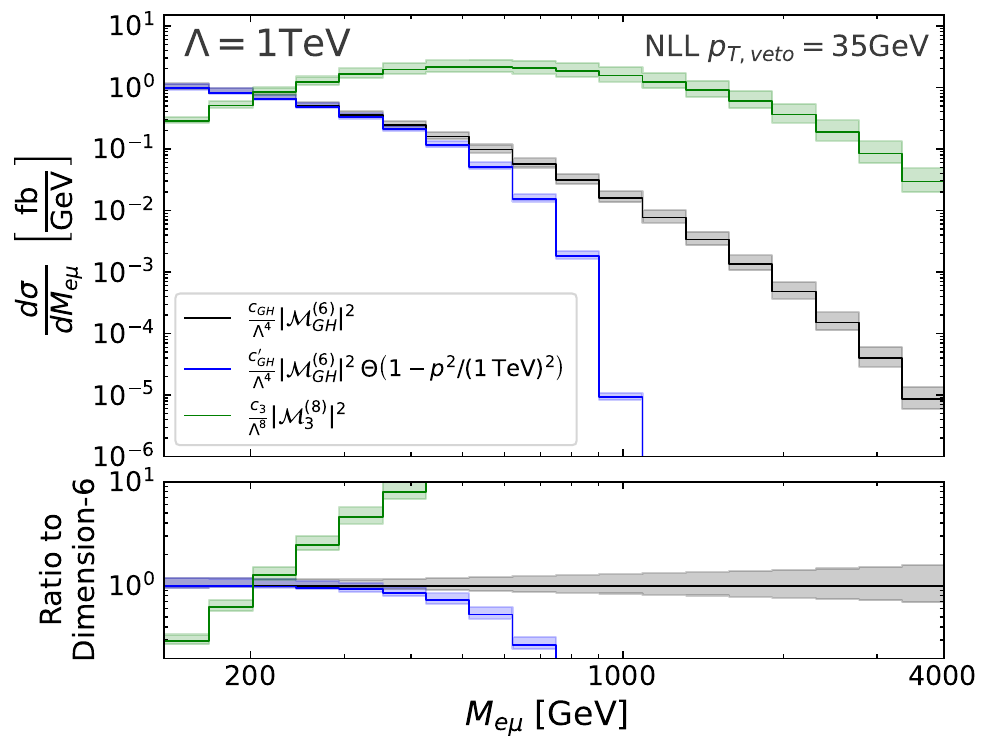}
    \includegraphics[width=.5\textwidth]{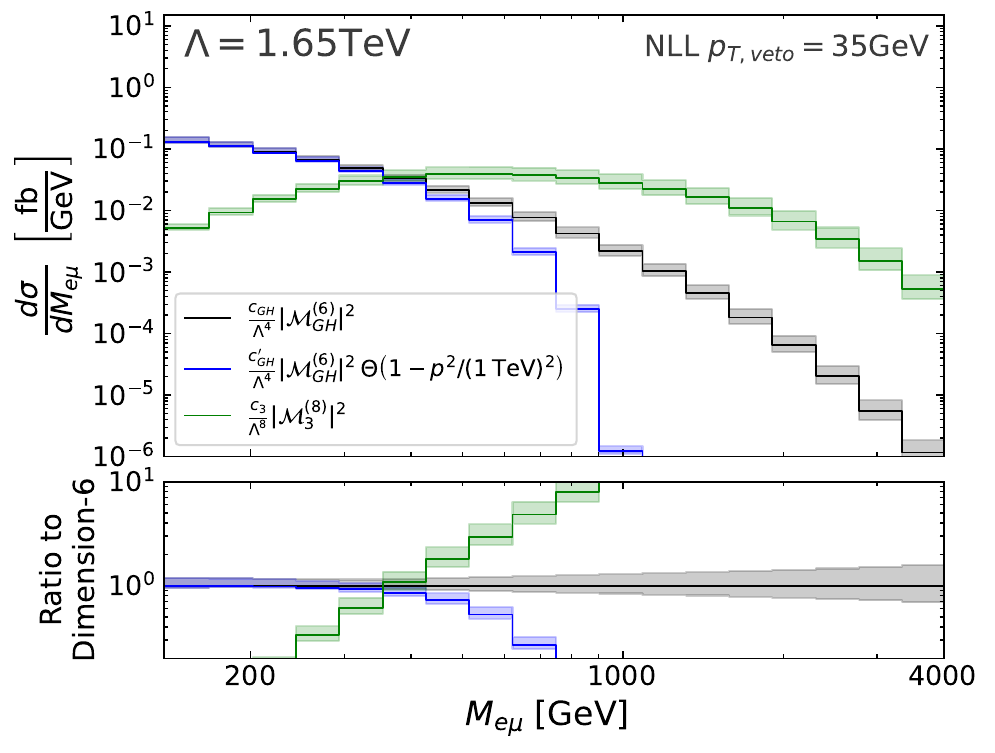}
    \includegraphics[width=.5\textwidth]{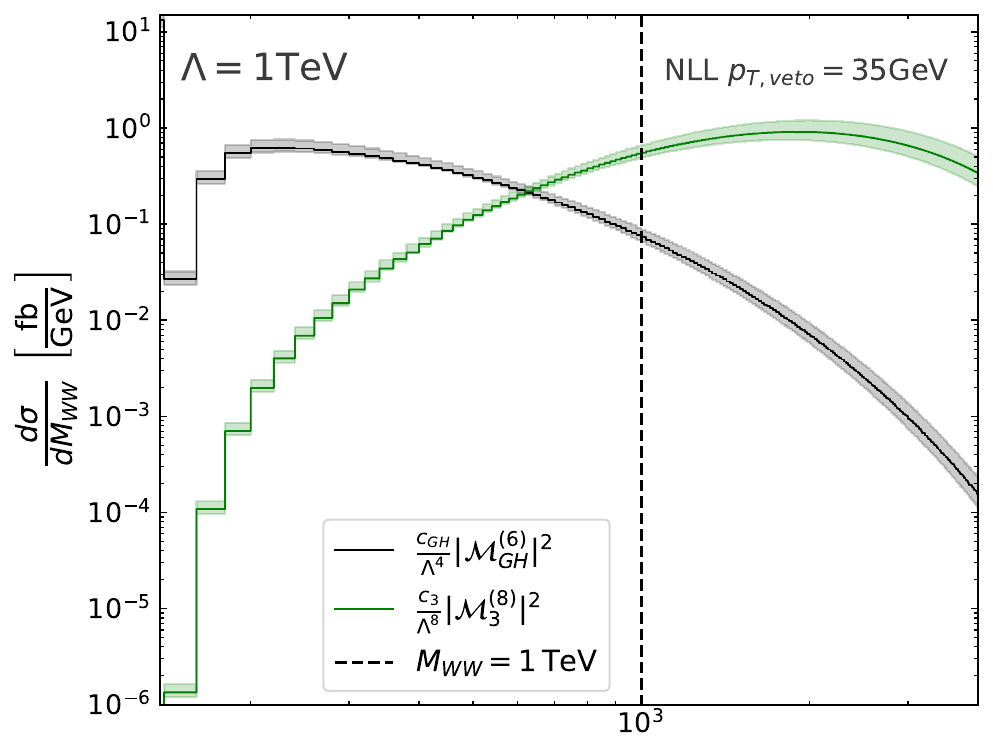} 
  \includegraphics[width=.5\textwidth]{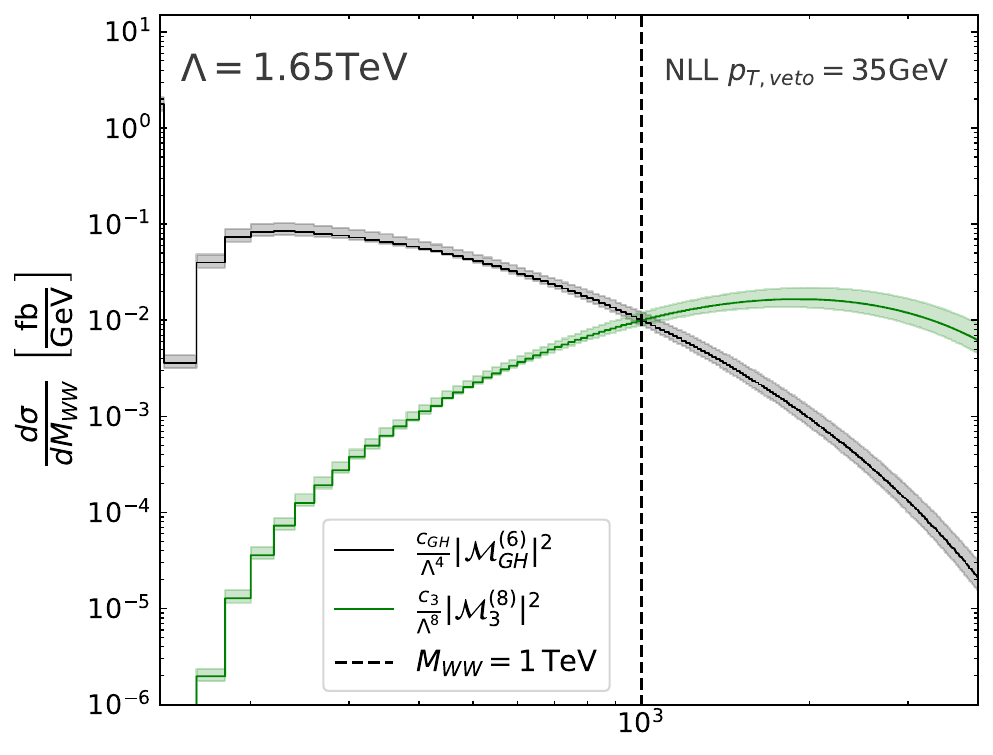} 
  \caption{Comparison of the full dimension-6 (black), dimension-6
    subject to the cut $M_{WW} < 1\,$TeV (blue), and dimension-8
    (green) contributions to the $M_{e\mu}$ (left) and $M_{WW}$
    (right) distributions.  The upper panels are for
    $\Lambda = 1\,$TeV, the lower panels for $\Lambda = 1.65\,$TeV.
  }
  \label{fig:EFTvalidcomparison}
\end{figure}

As anticipated in the introduction, the interference between
dimension-8 contributions and the SM is negligible. Therefore, we
concentrate here on the squared contributions of individual operators. These are
obtained from MCFM-RE and shown in
figure~\ref{fig:EFTvalidcomparison}.

We consider three classes of methods of ensuring this EFT validity as follows:

\begin{itemize}
  
\item {\bf Clipping on Simulation (CoS).}  By ``Clipping'' on the
  simulation, we refer to the method of enforcing $M_{WW} < \Lambda$
  on the EFT simulation alone, which is equivalent to fitting Wilson
  coefficients as
  $\frac{c^\prime_i}{\Lambda^2}\,\Theta(1 - p^2/\Lambda^2)$. This is
  achieved by setting a cut in the generation of the BSM
  prediction. However, since the cut is not actually placed on the
  data, the SM simulation is run as normal without the cut on
  $M_{WW}$.

\item {\bf Comparison bin-by-bin of dimension-6 and dimension-8
    operators (CBB).}  This method involves directly comparing the
  size of dimension-6 and dimension-8 operators.  Using the relation
  in eq.~(\ref{eq:empirical_lambdamin}), this is used to find the
  range of bins in a distribution where the dimension-8 squared
  contribution is negligible compared to the dimension-6 squared
  contribution. Assuming accurate modelling of both contributions,
  this method automatically satisfies our definition of ensuring that
  data used falls within the EFT valid regime.

\item {\bf Implementing a cut on data (CoD).}  The alternative method
  would be to implement a cut on the data itself which removes events
  with energies above the cut-off scale $\Lambda$.  Since $M_{WW}$ is
  not observable in this channel, a suitable proxy must be
  identified. $M_{e\mu}$ has already been shown to be a poor proxy for
  $M_{WW}$~\cite{Falkowski:2016cxu} but one could use other
  observables such as transverse-mass variables.
\end{itemize}

In figure~\ref{fig:EFTvalidcomparison}, we test the CoS method via
comparison with the CBB method, which we assume to be the most robust
way of ensuring EFT validity and has already been used
in~\cite{Gillies:2024mqp}.  When comparing to dimension-8 squared
contributions we always perform simulations without a cut. This is
because the cut does not appear in the data and is only a method to
reduce the size of dimension-6 contributions in bins where
dimension-8 dominates. From the upper left-hand panel of
figure~\ref{fig:EFTvalidcomparison}, it can be seen that imposing CoS
with $M_{WW}< 1\,$TeV for $\Lambda = 1\,$TeV does not remove
$M_{e\mu}$ bins where dimension-8 squared dominates
($M_{e\mu}>200\,$GeV). Specifically, the range
$200\,$GeV$< M_{e\mu} < 400\,$GeV still has a significant dimension-8
squared contribution even after CoS. This implies CoS is not a good
method for determining the range of EFT validity. There are two main
reasons for this. One is the fact the cut $M_{WW} < \Lambda$ is not
strict enough. As can be seen from the lower panels of
figure~\ref{fig:EFTvalidcomparison}, for a cut of $M_{WW}< 1\,$TeV to
be effective, $\Lambda $ must be at least $1.65\,$TeV. The second
reason is that, even when applying this stricter cut, the correlation
between $M_{e\mu}$ and $M_{WW}$ can change order by order in the EFT
expansion. In fact, as we add higher-order terms in the EFT
expansion, amplitudes grow more and more rapidly with energy. We
expect high $M_{WW}$ bins to overcome kinematical barriers and
contribute more and more to lower energy $M_{e\mu}$ bins. Therefore,
using dimension-6 predictions to estimate the contribution of higher
dimensional operators is not an effective strategy. This can be seen
in the upper right-hand panel of figure~\ref{fig:EFTvalidcomparison},
where CoS with $M_{WW}< 1.65\,$TeV for $\Lambda = 1\,$TeV does not
eliminate bins in the $M_{e\mu}$ distribution where dimension-8
squared dominates.

Although CBB is definitely more robust than CoS, it can be more cumbersome to implement in practice.
Firstly, it requires knowledge of the largest dimension-8 contributions. Second it requires re-evaluating
which bins can be used for every value of $\Lambda$. Therefore, it would be desirable to have a good measurable
proxy for $M_{WW}$ which implement a robust CoD procedure. This will be discussed in the following section.

\section{Transverse-Mass Observables}

To confirm the changing relationship between $M_{e\mu}$ and $M_{WW}$
(and that therefore $M_{e\mu}$ is not a good proxy for
$M_{WW}$), we compute $\mathbb{E}\left[M_{WW} \mid M_{e\mu}\right]$,
the conditional expectation value of $M_{WW}$ for a given $M_{e\mu}$ bin.  We
repeat this for the SM, dimension-6, and dimension-8 squared
contributions. Since the distribution of $M_{WW}$ for a given $M_{e\mu}$ bin is
skewed, we use the asymmetric variance to obtain an
error on $\mathbb{E}\left[M_{WW} \mid M_{e\mu}\right]$. 

\begin{figure}[!htbp]
\centering
 \includegraphics[width=.49\textwidth]{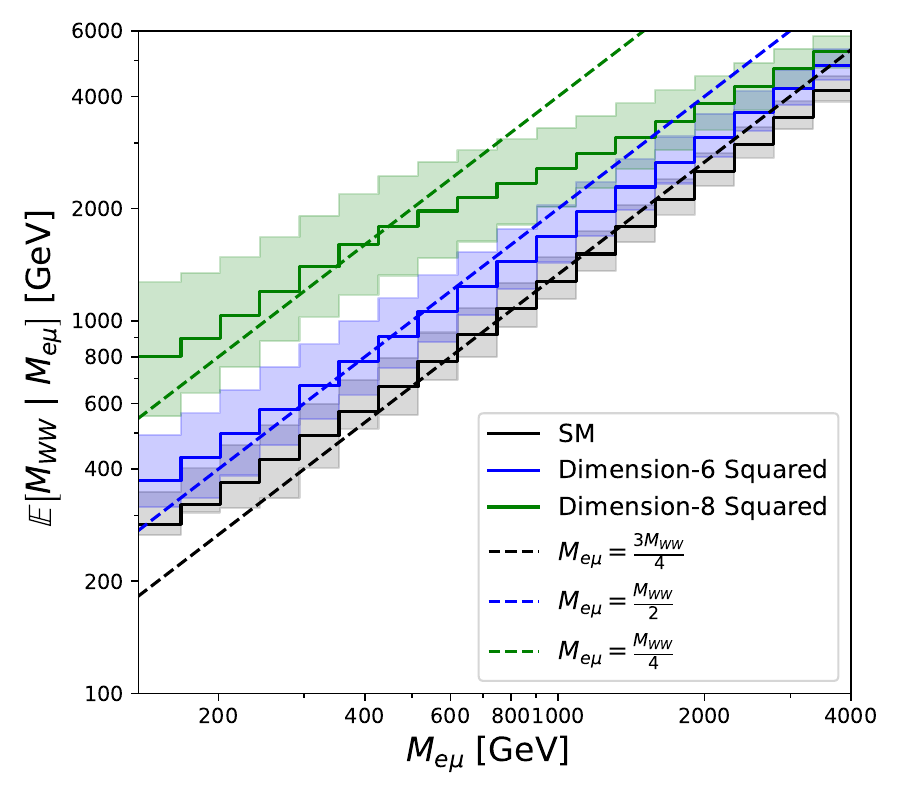}
\includegraphics[width=0.49\textwidth]{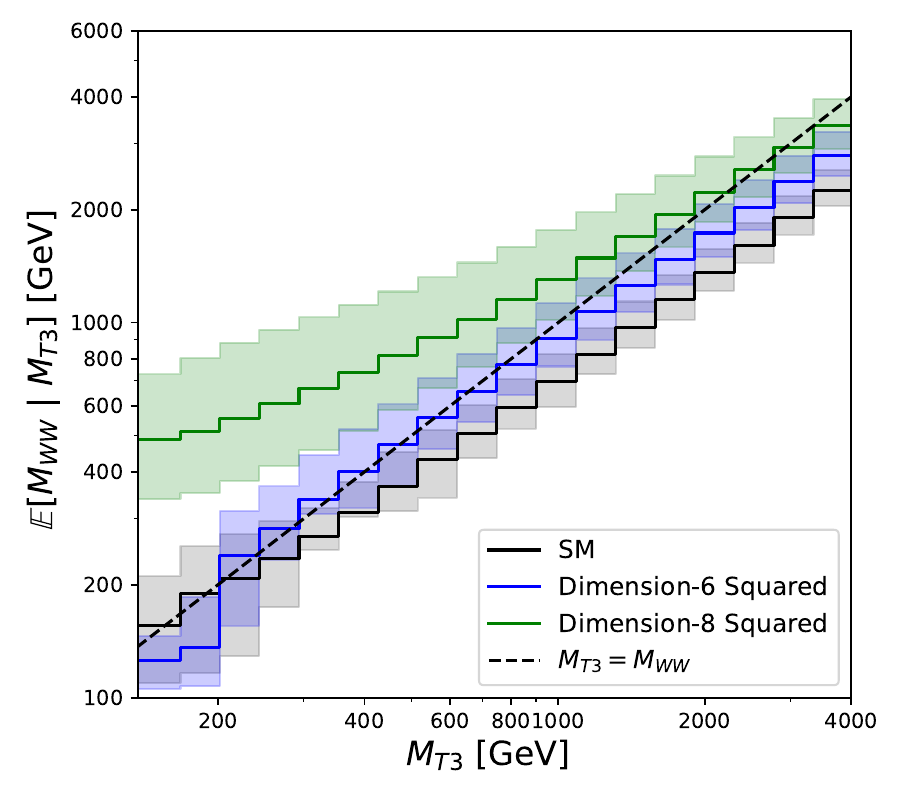}
 \includegraphics[width=0.49\textwidth]{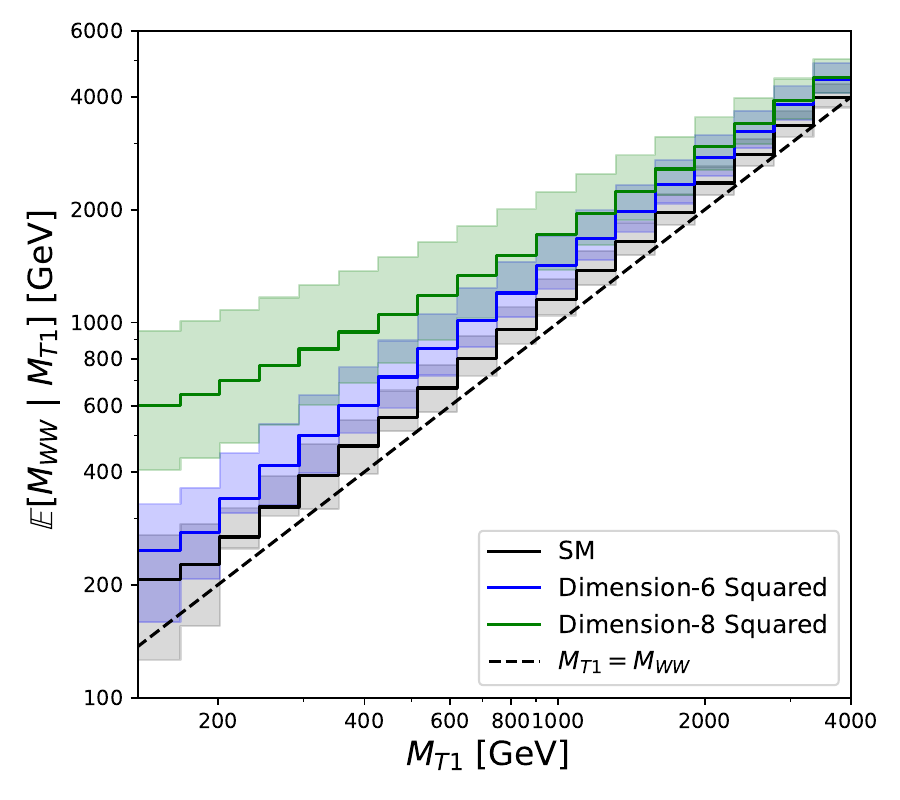}
\includegraphics[width=0.49\textwidth]{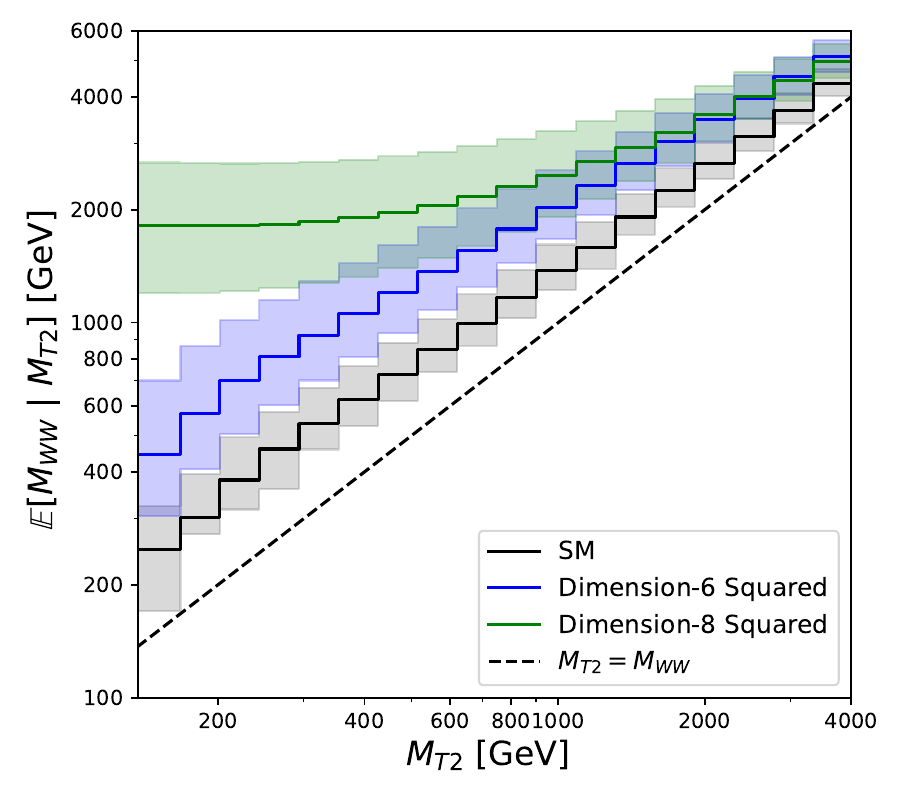}
\caption{The conditional expectation value of $M_{WW}$ for each
  bin of the $M_{e\mu}$ (top left),  $M_{T1}$ (bottom left), $M_{T2}$ (bottom right), and $M_{T3}$ (top right) distributions,
  for the SM (black), dimension-6 squared (blue),
  and dimension-8 squared (green) contributions. The error is given by 
  calculating the asymmetric variance from the expectation value.}
  \label{fig:correlation_mtall}
\end{figure}

The conditional expectation value
$\mathbb{E}\left[M_{WW} \mid M_{e\mu}\right]$ (along with an error
from its asymmetric variance) is shown in the upper-left panel of
figure~\ref{fig:correlation_mtall} at NLL accuracy for the SM $gg$
contribution, the dimension-6 ($\phi^\dagger\phi\,GG$) and dimension-8
($WWGG$) squared contributions. The expected $M_{WW}$ is plotted for
$18$ logarithmically spaced bins from $138 < M_{e\mu} < 4000\,$GeV.
For all the expectation value plots in this section, we use the
``NNPDF40$\_$nnlo$\_$as$\_$01180$\_$nf$\_$4'' PDF
set~\cite{NNPDF:2021njg}.

In the SM, there is a strong correlation between $M_{e\mu}$ and
$M_{WW}$ with $M_{e\mu}\approx \frac{3M_{WW}}{4}$. In fact, the
SM cross-section falls rapidly with energy, both due to phase-space
constraints and the effect of the jet veto.  Whilst we might na\"ively
expect on average $M_{e\mu} \approx \frac{M_{WW}}{2}$, the rapidly
decaying nature of the cross-section means that lower energy $M_{WW}$
bins will contribute more, which skews the average closer to
$M_{e\mu} = M_{WW}$. At dimension-6, there is still some correlation
between $M_{e\mu}$ and $M_{WW}$. However, the relation between the two
changes as the dimension-6 contribution no longer decreases as rapidly
with energy compared to the SM. As a consequence, we now get much
closer to the na\"ive expectation $M_{e\mu} \approx
\frac{M_{WW}}{2}$. At dimension-8, there is a much greater variance,
showing that the correlation between $M_{WW}$ and $M_{e\mu}$ is
weaker. Furthermore, given an $M_{e\mu}$ bin, the events which
contribute have a much higher $M_{WW}$ value for dimension-8. In this
case, the relation at lower $M_{e\mu}$ is
$M_{e\mu} \approx \frac{M_{WW}}{4}$ for lower $M_{e\mu}$ values.  Note
that, for all distributions,
$\mathbb{E}\left[M_{WW} \mid M_{e\mu}\right]$ converges to
$M_{e\mu} = M_{WW}$ at the $14\,$TeV kinematical barrier.


To summarise, the dimension-8 correlation differs quite substantially
from that of the dimension-6 contribution. Lower $M_{e\mu}$
bins do not get as much contribution from high-$M_{WW}$ events in the
dimension-6 case when compared to the dimension-8 case.
The fact that $M_{e\mu}$ and $M_{WW}$ correlate poorly means that the
EFT breakdown happens at very low energies for the $M_{e\mu}$
observable. With this in mind, we try to find better observables which
could more closely track $M_{WW}$, which we remind is not observable for leptonic $W$ decays.

\begin{figure}[!htbp]
\centering
  \includegraphics[width=0.51\textwidth]{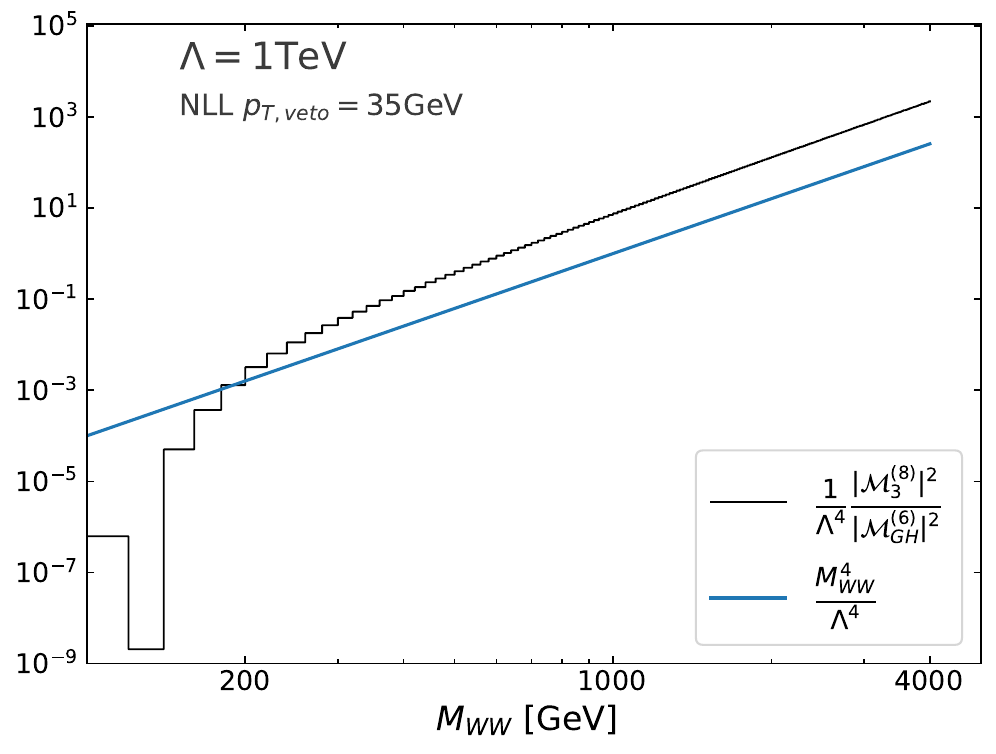}
  \includegraphics[width=0.49\textwidth]{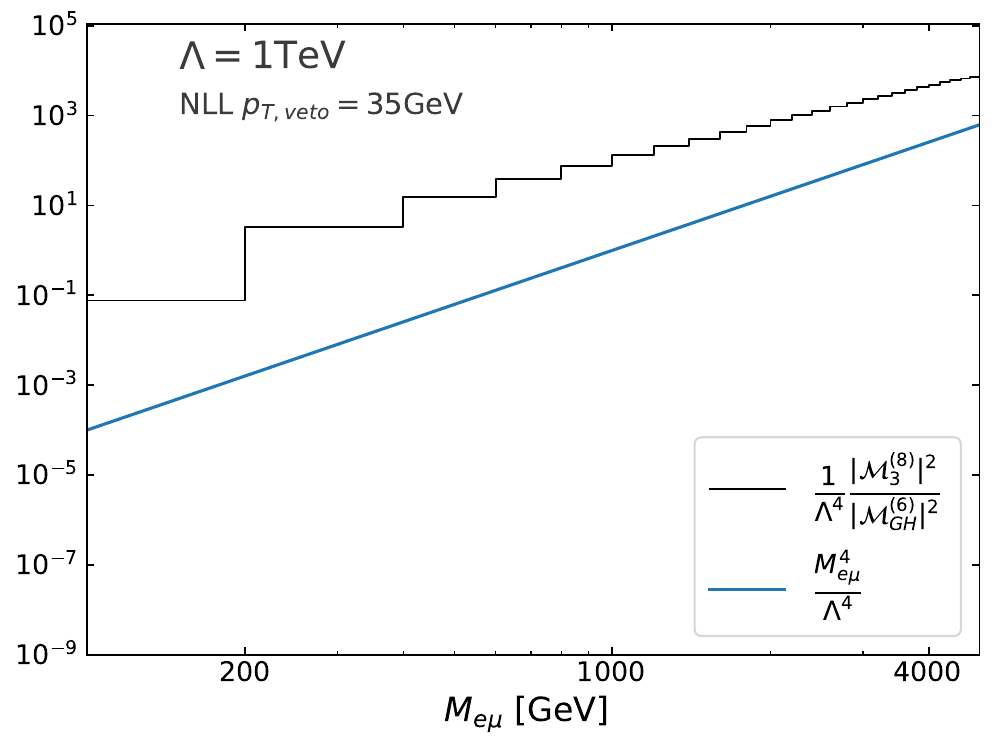}
  \includegraphics[width=0.49\textwidth]{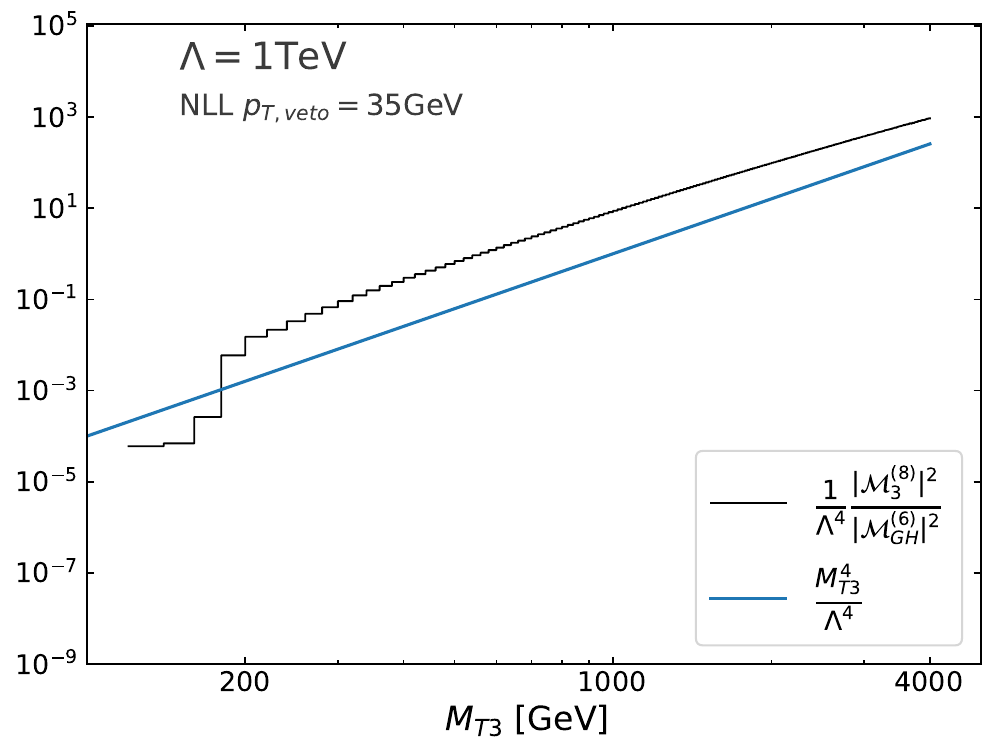}
  \includegraphics[width=0.49\textwidth]{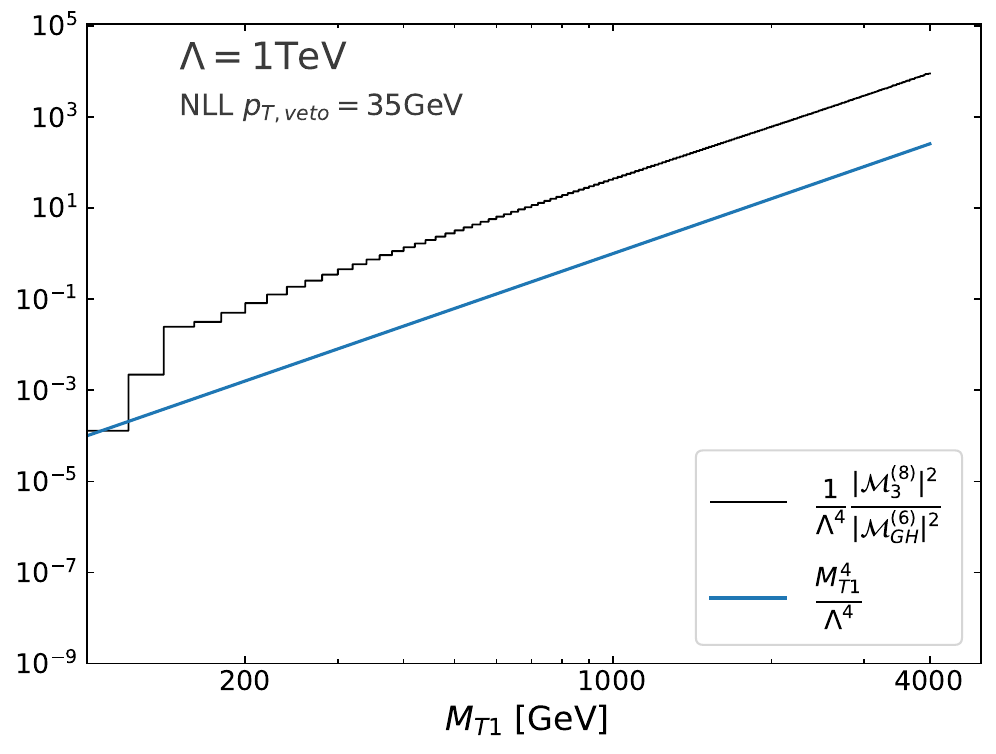}
  \includegraphics[width=0.49\textwidth]{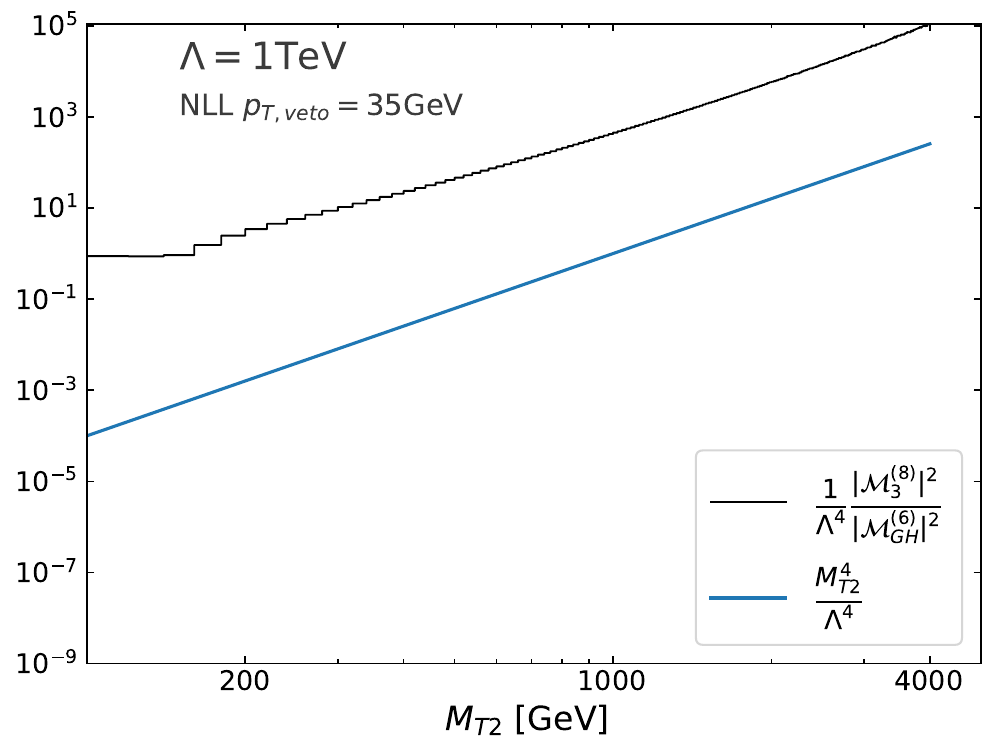}
  \caption{Comparison of the ratio of dimension-8 squared and dimension-6
  squared contributions to the theoretical value from the EFT (for the $M_{WW}$
  distribution). This theoretical value is given as $\frac{M^4}{\Lambda^4}$, $M=\{M_{WW}
  ,\,M_{T1},\,M_{T2},\,M_{T3},\,M_{e\mu}\}$.}
  \label{fig:MWWMT_rat}
\end{figure}

In the literature~\cite{Arpino:2019fmo}, we found three transverse-mass observables which
were designed to act as a good proxy of the $M_{WW}$
distribution. These are defined as:
\begin{subequations}\label{eq:MTdefs}
\begin{align}
M_{T1} &\equiv \sqrt{ \left( M_{T,e\mu} + p_T\!\!\!\!\!\!/ \>\right)^2 - \left( \vec{p}_{T,e\mu} + \vec{p}_T\!\!\!\!\!\!/ \>\> \right)^2 }, \\
M_{T2} &\equiv \sqrt{ 2 p_{T,e\mu}\, p_T\!\!\!\!\!\!/ \>\> \big(1 - \cos \Delta \phi_{e\mu,\text{miss}}\big) }, \\
M_{T3} &\equiv \sqrt{ \big( M_{T,e\mu} + M_T\!\!\!\!\!\!\!/ \>\> \big)^2 - \big( \vec{p}_{T,e\mu} + \vec{p}_T \!\!\!\!\!\!/ \>\> \big)^2 },
\end{align}
\end{subequations}
with
\begin{equation}\label{eq:MTdefs2}
M_{T,e\mu} \equiv \sqrt{ p_{T,e\mu}^2 + M_{e\mu}^2 }\,,\qquad
M_T\!\!\!\!\!\!\!/ \>\> \equiv \sqrt{ p_T^{\,2}\!\!\!\!\!\!/ \>\> + M_{e\mu}^2 }.
\end{equation}
The conditional expectation values of $M_{WW}$ for the given
observables for the SM and at dimension-6 and dimension-8
are shown in figure~\ref{fig:correlation_mtall}.
Of the four observables, the one which correlates best with $M_{WW}$
is $M_{T3}$. We can also appreciate this fact by considering how the ratio between
dimension-6 and dimension-8 compares with $M^4/\Lambda^4$
($M=M_{T1},M_{T2},M_{T3},M_{WW}$), which should be of the order of
magnitude of the ratio between dimension-6 and dimension-8,
particularly for $M=M_{WW}$. We know this because the dimension-6 and dimension-8 squared 
contributions grow as $M_{WW}^4/\Lambda^4$ and $M_{WW}^8/\Lambda^8$ respectively. This is shown in
figure~\ref{fig:MWWMT_rat}. It can be seen that $M_{T3}$ behaves the
most closely to $M_{WW}$.
In particular, it
can be seen from figure~\ref{fig:MWWMT_rat}, that $M_{e\mu}$ performs
markedly worse as a proxy for $M_{WW}$. 

\begin{figure}[!htbp]
\centering
  \includegraphics[width=0.495\textwidth]{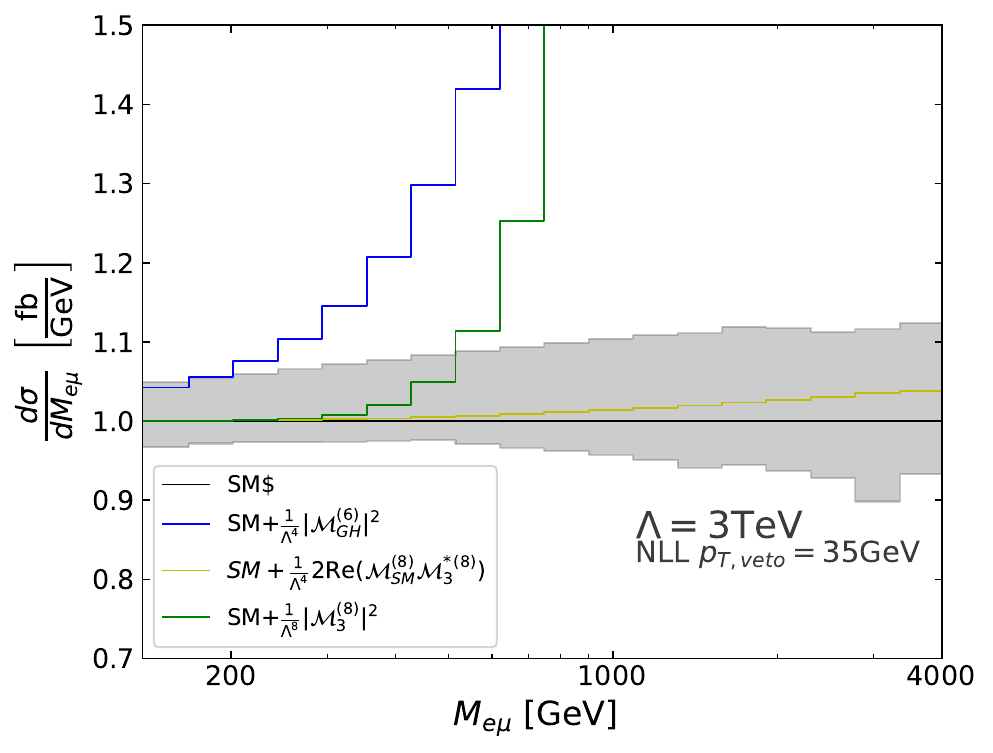}
  \includegraphics[width=0.495\textwidth]{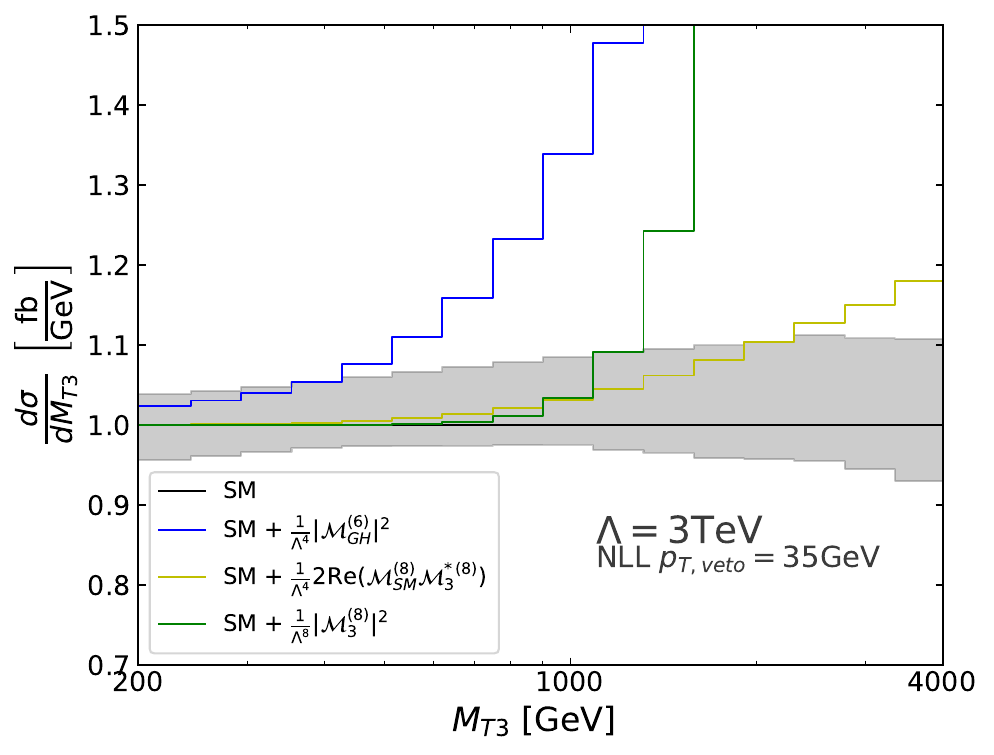}
  \caption{Comparison of $M_{T3}$ and $M_{e\mu}$ distributions of SM
    (black) with SM theoretical uncertainty (shaded area), dimension-6
    squared contribution (blue) and dimension-8 squared (green) and SM
    interference (yellow) contributions, for $\Lambda=3\,$TeV.}
  \label{fig:mt3_memu_comparison} 
\end{figure}

\newpage
To have an indication of the values of $M_{e\mu}$ and $M_{T3}$ for
which the EFT breaks down at $\Lambda=3\,$TeV, in
figure~\ref{fig:mt3_memu_comparison} we compare the relative sizes of
dimension-6 quadratic, dimension-8 linear, and dimension-8 quadratic
contributions for the $M_{e\mu}$ and $M_{T3}$ distributions .  We can
see that the dimension-8 interference is still negligible in the
region of validity of the EFT expansion for both $M_{T3}$ and for
$M_{e\mu}$. It can also be seen that the $M_{T3}$ observable gives a
larger range of bins where the EFT is valid for a given
$\Lambda$. Also, since $M_{T3}$ correlates so much better with
$M_{WW}$, the Higgs peak appears at around the value of
$M_{T3} = 125\,$GeV, whereas for $M_{e\mu}$ it appears at energies
$M_{e\mu} < 100\,$GeV. In the following, we will focus only on the
tails of the distributions, excluding any effect from the Higgs peak,
and we only look at bins for $M_{T3} > 200\,$GeV and
$M_{e\mu} > 138\,$GeV.

\section{Sensitivity Studies}

\begin{figure}[!htbp]
  \includegraphics[width=.5\textwidth]{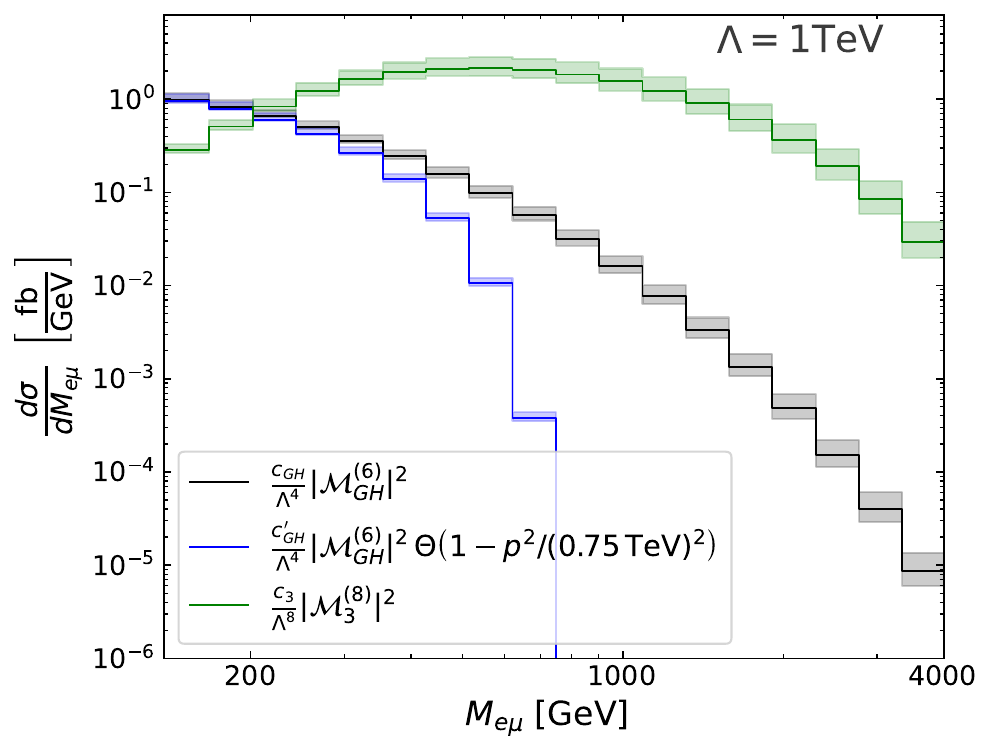}
  \includegraphics[width=.5\textwidth]{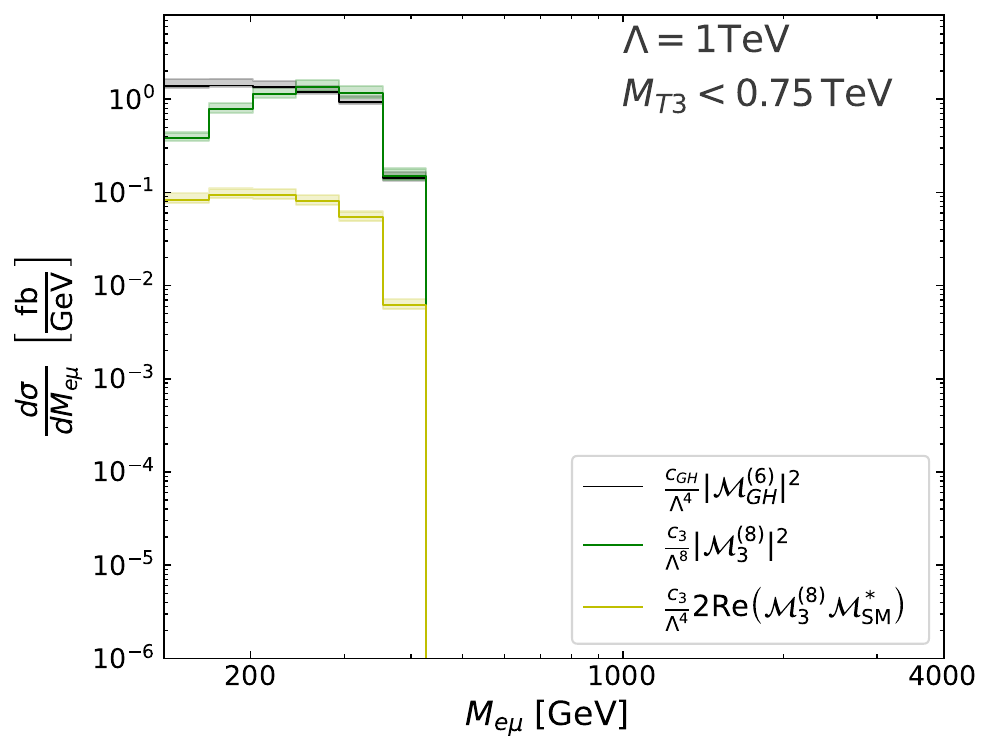}
  \caption{Comparison of the cut on simulation (CoS) approach (left) with the cut on data (CoD)
  approach (right). The CoD approach includes an extra fiducial cut of $M_{T3} < 0.75\,$TeV whereas the CoS
  approach applies a cut of $M_{WW} < 0.75\,$TeV only to the generation of dimension-6 EFT contributions. 
  The dimension-6 contribution without CoS is shown in black. The dimension-6
  contribution with CoS is shown in blue. Dimension-8
    squared (green) and interference (yellow) contributions are also shown without a CoS.}
  \label{fig:EFTvalidcomparisonmt3}
\end{figure}

We can set constraints on new physics by using the cut on data (CoD)
method outlined in the previous using $M_{T3}$ as a proxy for $M_{WW}$. This ensures EFT validity for
these constraints without needing to profile the size of dimension-8 operators. The cut can
either be applied directly onto the $M_{T3}$
distribution, or can be used with another distribution (e.g. in $M_{e\mu}$), by ensuring that 
the $M_{T3}$ cut is added to the fiducial cuts of table~\ref{tab:fiducial-cuts}. 
To compare the CoD method with the $M_{e\mu}$ distribution to what is in use by ATLAS (CoS),
in figure~\ref{fig:EFTvalidcomparisonmt3} we plot the $M_{e\mu}$
distribution with the cut on data of $M_{T3} < 750\,$GeV and
the cut on simulation of dimension-6 at $M_{WW} < 750\,$GeV for an EFT scale $\Lambda=1\,$TeV. In fact, using
the $M_{WW}$ distribution, we found that a more appropriate value for
the CoD or CoS would have been at most $650\,$GeV. However, applying the 
    cut $M_{T3} < 750\,$GeV still keeps dimension-8 at or around the same level
    as dimension-6 for the CoD approach. 

We then performed sensitivity studies to compare the method of CoS on the
dimension-6 EFT simulation (at $M_{WW} < 750\,$GeV), to the method of
implementing a CoD at $M_{T3} < 750\,$GeV.  We compared both of these
to the method of ensuring that dimension-6 is larger than dimension-8
bin by bin (CBB).  We also investigated if we could obtain better
constraints using the $M_{T3}$ distribution when compared to
constraints with the $M_{e\mu}$ distribution, both with the CBB
method.  We used pure NNLL predictions for the SM $q\bar q$ channel as this
allowed us to use MCFM-RE, which was faster than computing the full
NNLO+NNLL prediction with MATRIX+RadISH. In~\cite{Gillies:2024mqp}, we
showed that the agreement between the two predictions is within $5\%$,
and with even better agreement at high values of $M_{e\mu}$ and
$M_{WW}$. Since the SM $gg$ contribution is heavily suppressed
($<5\%$ of the $q\bar q$ contribution~\cite{Gillies:2024mqp}) at high
energies, we neglect it for this study.  This is true for both the
$M_{e\mu}$ and $M_{WW}$ distributions, the latter of which should be
correlated to the transverse mass observable $M_{T3}$. We do not take
into consideration EW corrections which make these contour plots more
conservative (as EW corrections give a large Sudakov suppression in
the tails of the SM distribution).

For our sensitivity studies, we use the method
of~\cite{Gillies:2024mqp}. Using eq.~\eqref{eq:Mgg} and our BSM
prediction at NLL we can define, for a set of $c_i$ and $\Lambda$, a
prediction at the HL-LHC which we call
$\{m_j\left(c_i, \Lambda\right)\}$. We can then compare this to data
points $\{n_j\}$ obtained by drawing a Poisson distribution from the
NNLL SM predictions. To ensure EFT validity, we only take
$\{n_j\}$ bins, up to the largest bin $N$ which satisfies
eq.~\eqref{eq:empirical_lambdamin} for the given value of $\Lambda$.

For the generation of exclusion plots and sensitivity studies we then use a delta chi-squared test statistic defined as:
\begin{equation}
  \label{eq:deltachisq}
    \Delta \chi^2\left(c_i, \Lambda\right) \equiv \chi^2\left(c_i, \Lambda\right) - \chi^2\left(\hat c_i, \hat \Lambda\right)\,,
\end{equation}
where $\chi^2\left(c_i, \Lambda\right)$ is defined as:
\begin{equation}
  \label{eq:chisq}
    \chi^2\left(c_i, \Lambda\right) \equiv \sum_{j=1}^N \frac{\left(n_j - m_j\left(c_i, \Lambda\right)\right)^2}{\left(\Delta m_j\left(c_i, \Lambda\right)\right)^2}\,,
\end{equation}
and $ \hat c_i $ and $ \hat \Lambda $ 
are values which
minimise $ \chi^2\left(c_i, \Lambda\right) $. For each value of $N$, the
$\hat c_i$ and $\hat \Lambda$ must be found separately.
In order to account for
theoretical and systematic errors, following~\cite{Arpino:2019fmo}, we use
\begin{equation}
  \label{eq:Dni-exp}
  (\Delta m_j(c_i, \Lambda))^2 = m_j\left(c_i, \Lambda\right) + (\Delta^{(\mathrm{th})}_j/2)^2  + (\Delta^{(\mathrm{sys})}_j/2)^2 \,.
\end{equation}
In the above equation, $\Delta^{(\mathrm{th})}_j$ is the theoretical scale variation
uncertainty associated with the BSM prediction $m_j$.
The quantity $\Delta^{(\mathrm{sys})}_j$ gives an expected experimental systematic
error at the HL-LHC, computed by extrapolating
current systematic errors from ATLAS~\cite{ATLAS:2019rob} to higher energies.

To obtain constraints for the HL-LHC sensitivity studies we use the
method of median significance. This is done by generating many sets of
$\{n_j\}$ using the expected $\{\bar n_j\}$ given by the Standard
Model best prediction and a Poisson distribution for each bin
independently. For these simulated data sets we obtain the probability
distribution for $\Delta\chi^2\left(\{c_i, \Lambda\}\right)$, whose median
makes it possible to calculate the p-value associated with each
considered $\{c_i, \Lambda\}$. We then exclude all values of
$\{c_i, \Lambda\}$ whose p-value is less than 0.05. We then present
results obtained by using $18$ logarithmically spaced bins in the
$M_{e\mu}$ distribution and the $16$ logarithmically spaced bins in
the $M_{T3}$ distribution.

We note that when performing EFT constraints, we can only fit the combined
factor $c_i/\Lambda^n$ (with $n$ determined by the dimension of the operator).
There are two options for extracting constraints which is either to set $c_i = 1$
or fix $\Lambda$ and extract constraints on $c_i$. The ideal scenario is to fit both 
simultaneously, with $\Lambda$ determining the number of bins used in each fit 
for $c_i$ giving full multidimensional constraints. However, for the purposes of comparison
of methods, we fix $\Lambda = 1\,$TeV and extract constraints on $c_i$.

\begin{figure}[!htbp]
\centering
  \includegraphics[width=0.4\textwidth]{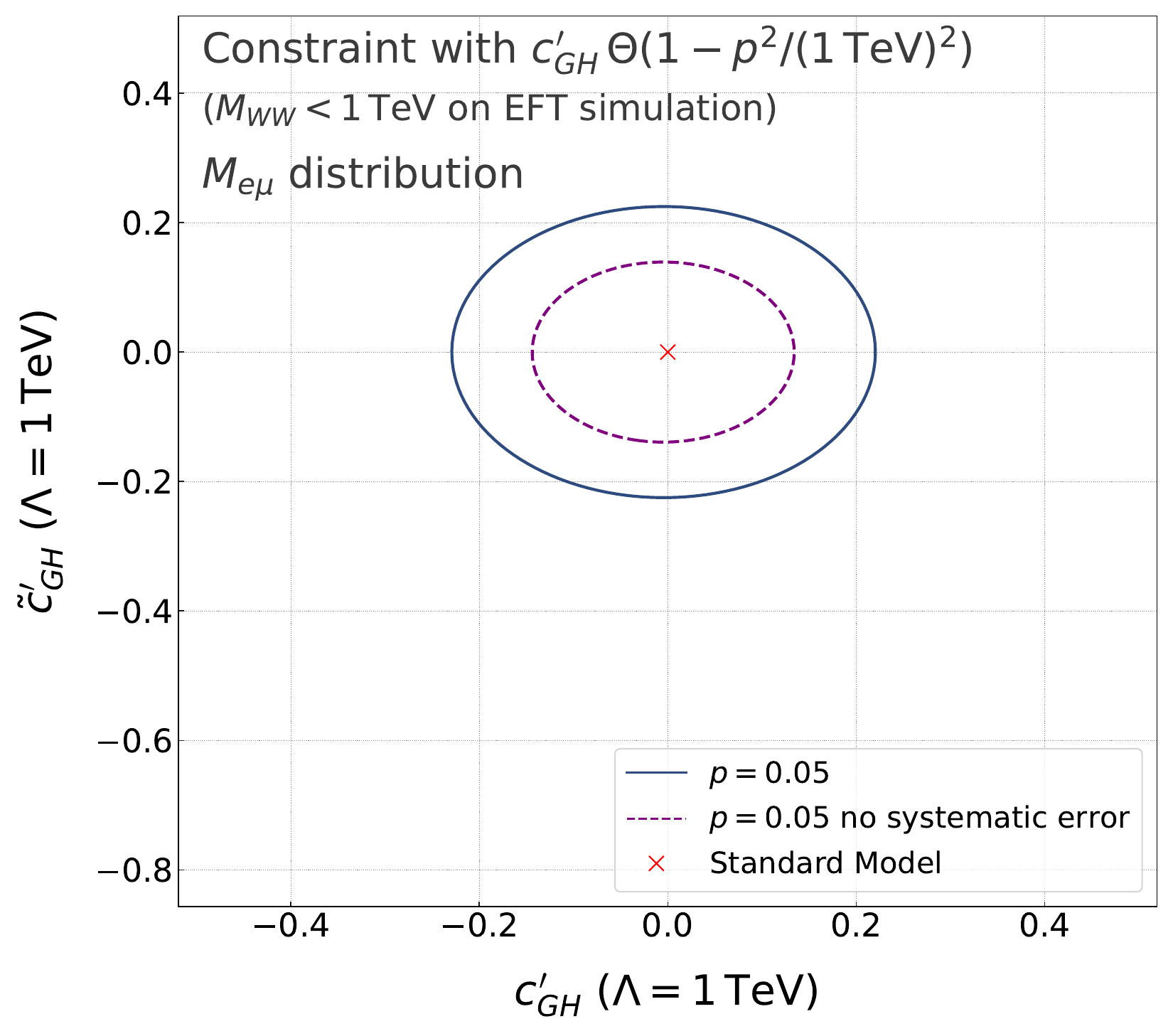}
  \includegraphics[width=0.4\textwidth]{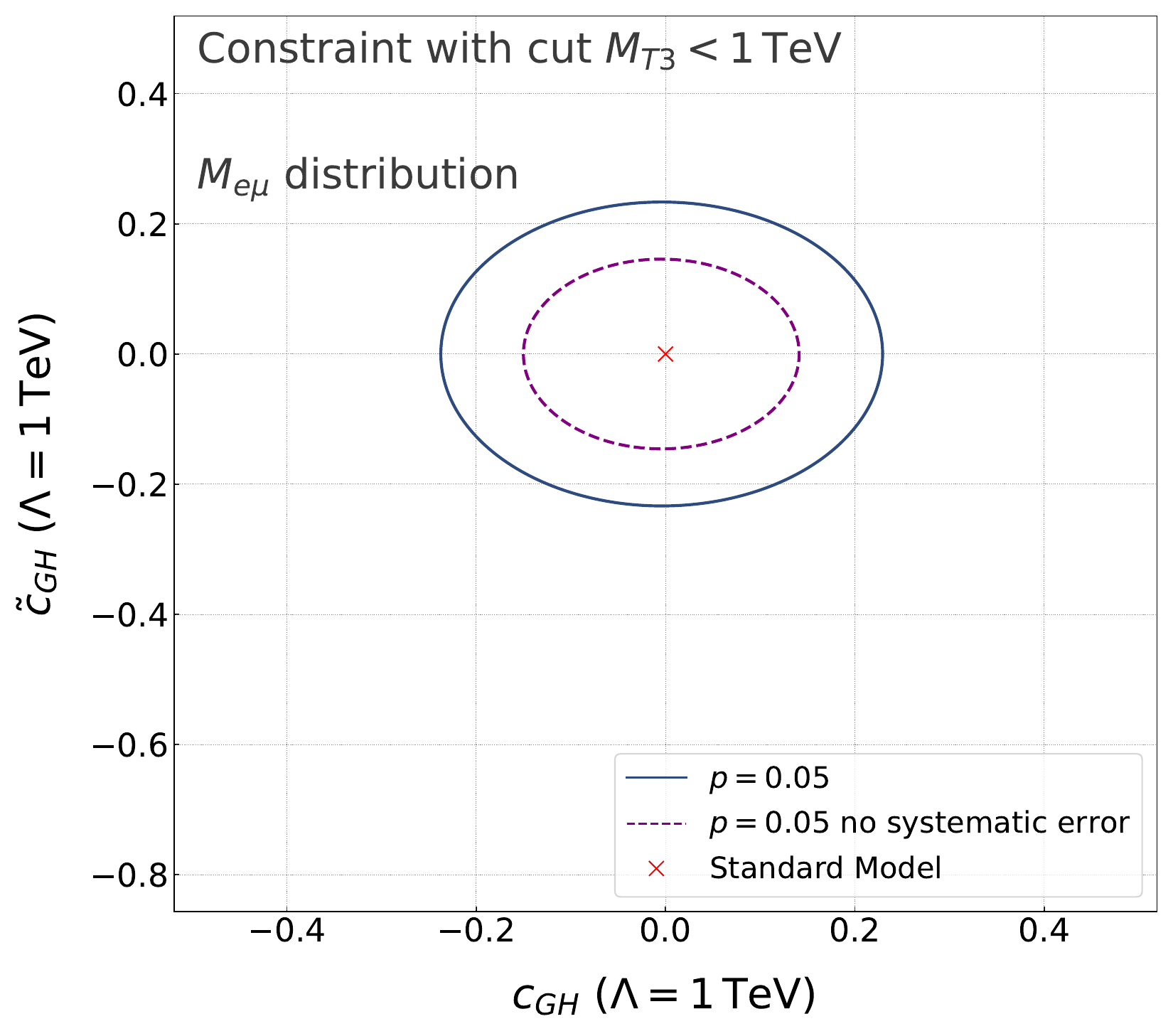}
    \includegraphics[width=0.4\textwidth]{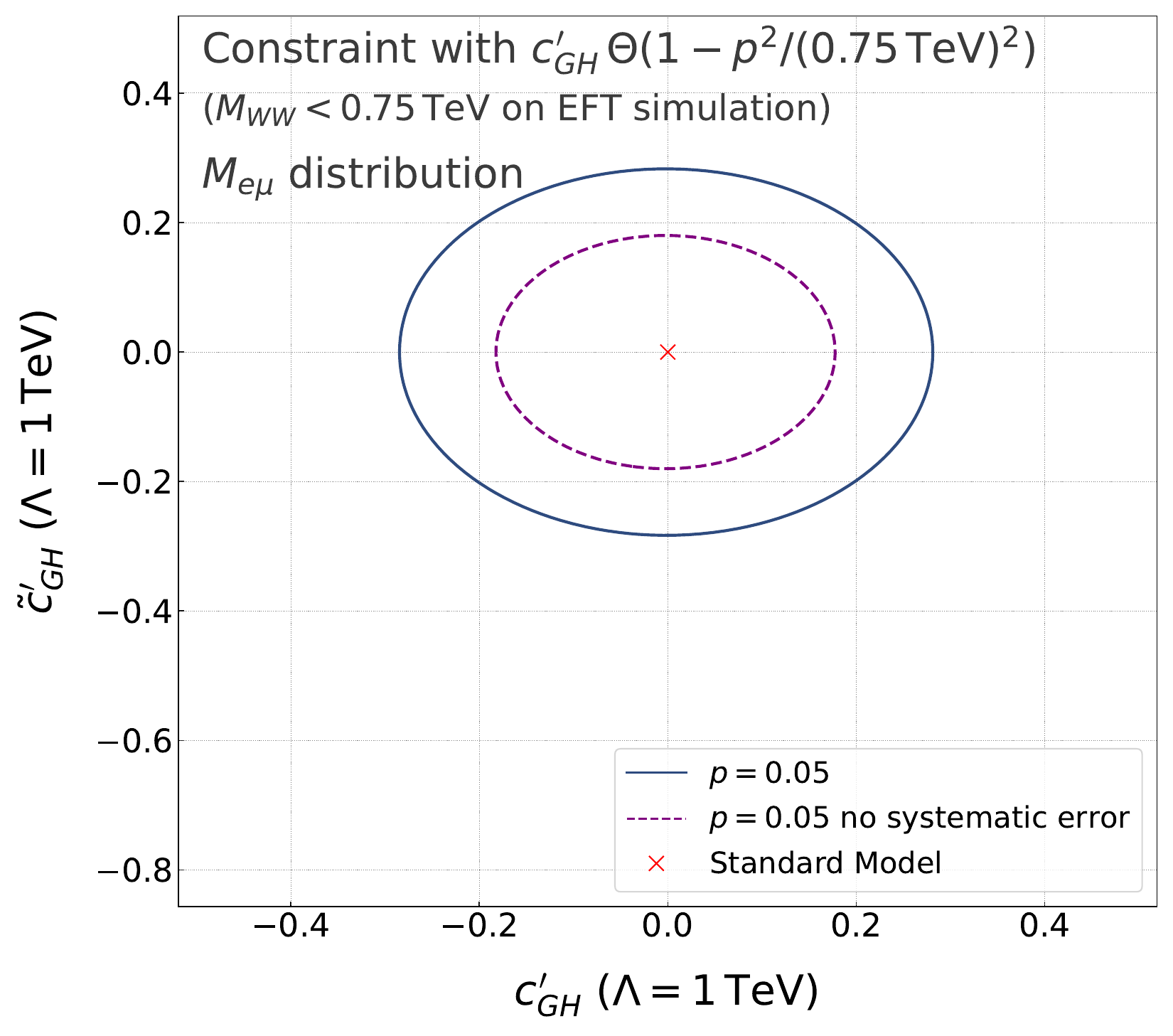}
  \includegraphics[width=0.4\textwidth]{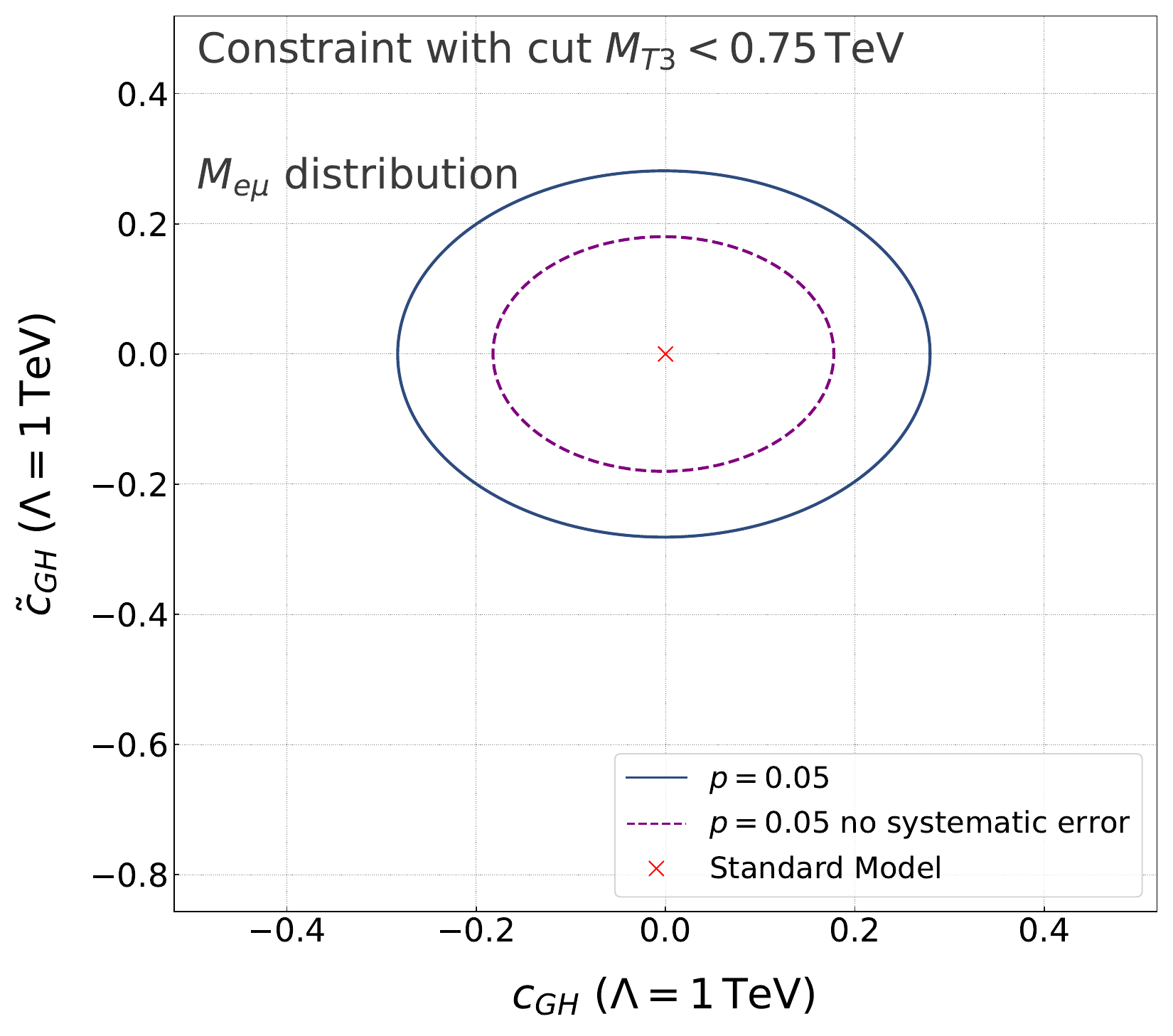}
  \caption{Sensitivity studies on the CP-even and CP-odd version of the 
  dimension-6 operator ($\phi\phi\, GG$) performed on projected HL-LHC data.
  On the right, an experimental cut has been placed at $M_{T3} < \{1\,(\mathrm{top}),\, 
  0.75\,(\mathrm{bottom})\}$. On the left, the simulation has been clipped for the 
  dimension-6 only such that $M_{WW} <\{1\,(\mathrm{top}),\, 
  0.75\,(\mathrm{bottom})\}$ which is equivalent to modifying the SMEFT coefficient
  by a Heaviside function with transition at the relevant scale. The constraints
  are approximately the same between the two methods with a slightly stronger 
  constraint for $M_{WW} <0.75\,$TeV than for the cut $M_{T3} <0.75\,$TeV.}
  \label{fig:contour_plot_comparisonMWW}
\end{figure}

First, we study clipping only the EFT simulation (CoS) (with no high
energy cut on the data).  We use the cuts $M_{WW} < \Lambda$ and
$M_{WW} < 0.75\,\Lambda$ (equivalent to replacing $c_i$ with
$c^\prime_i \Theta\left(1 - p^2 / (f\Lambda^2)\right)$ with
$f = \{0.75,\, 1\}$ respectively).  We compare this procedure to using
a cut of $M_{T3} < \Lambda$ and $M_{T3} < 0.75\,\Lambda$ directly on
the data (CoD), and fitting the $M_{e\mu}$ distribution with those
cuts. The corresponding contour plots are shown in
figure~\ref{fig:contour_plot_comparisonMWW}.  Although the CoD method
is more conservative, it gives comparable bounds with respect to
CoS. This is partly due to the fact that CoD acts on the SM background
as well, thus increasing the relative size of BSM contributions. Also,
the Wilson coefficients extracted with CoD have a clear interpretation
in terms of a local EFT, which is not the case for CoS.

We can also compare CoS and CoD to the CBB method, i.e.\ setting
$\Lambda = 1\,$TeV, and using eq.~\eqref{eq:empirical_lambdamin} to
determine the EFT valid bins. The contours obtained with CBB applied
to the $M_{e\mu}$ distribution are shown in the left panel of
figure~\ref{fig:contour_plot_comparisonM_{T3}}. It can be noted that
CBB gives a more stringent condition on data than CoS and CoD with the
current setup, leading to looser constraints on the Wilson
coefficients. More stringent cuts on $M_{WW}$ (for CoS) and $M_{T3}$
(for CoD), e.g. $M_{WW}/M_{T3}<650\,$GeV lead to similar constraints.
\begin{figure}[!htbp]
\centering
  \includegraphics[width=0.4\textwidth]{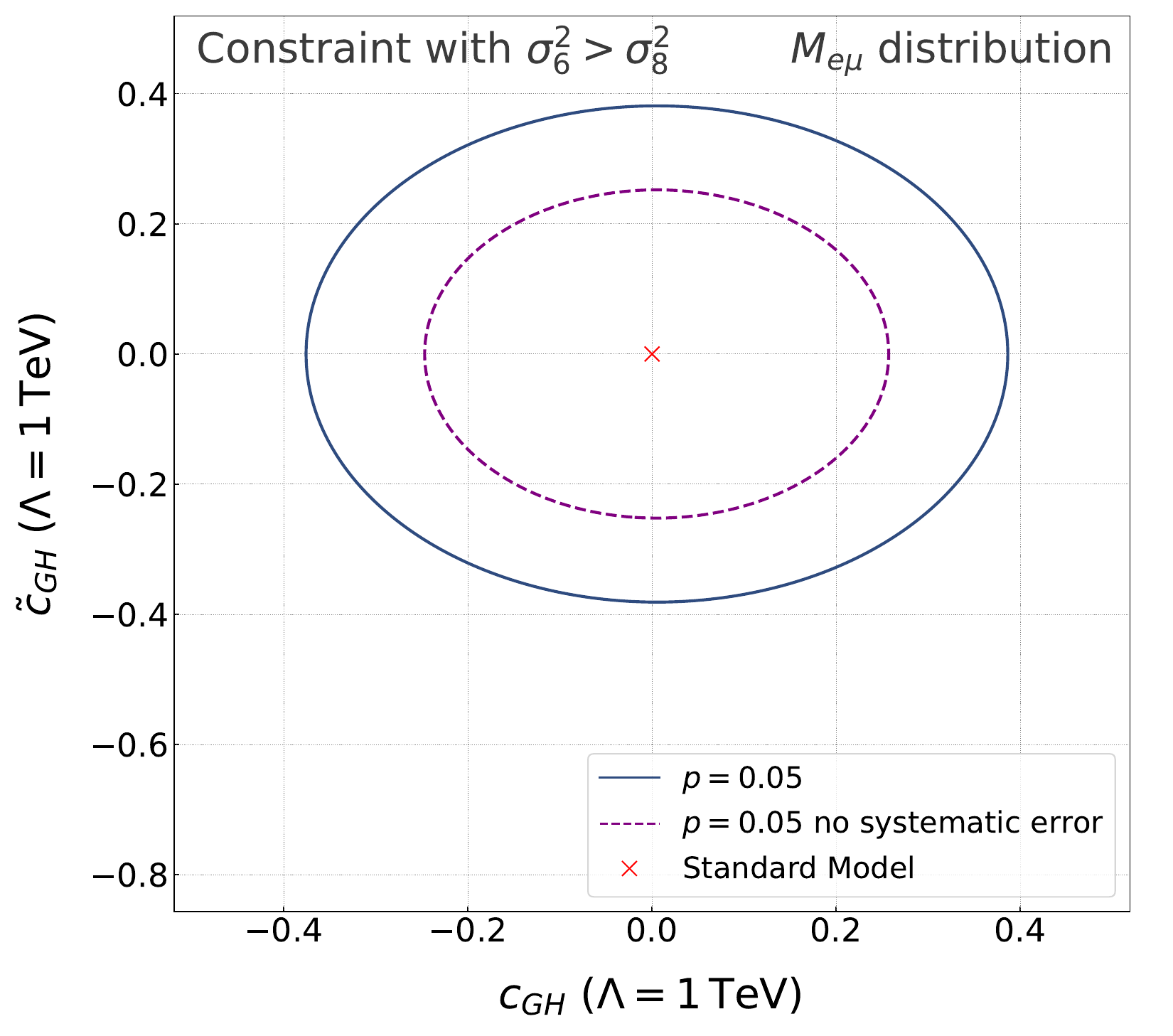}
  \includegraphics[width=0.4\textwidth]{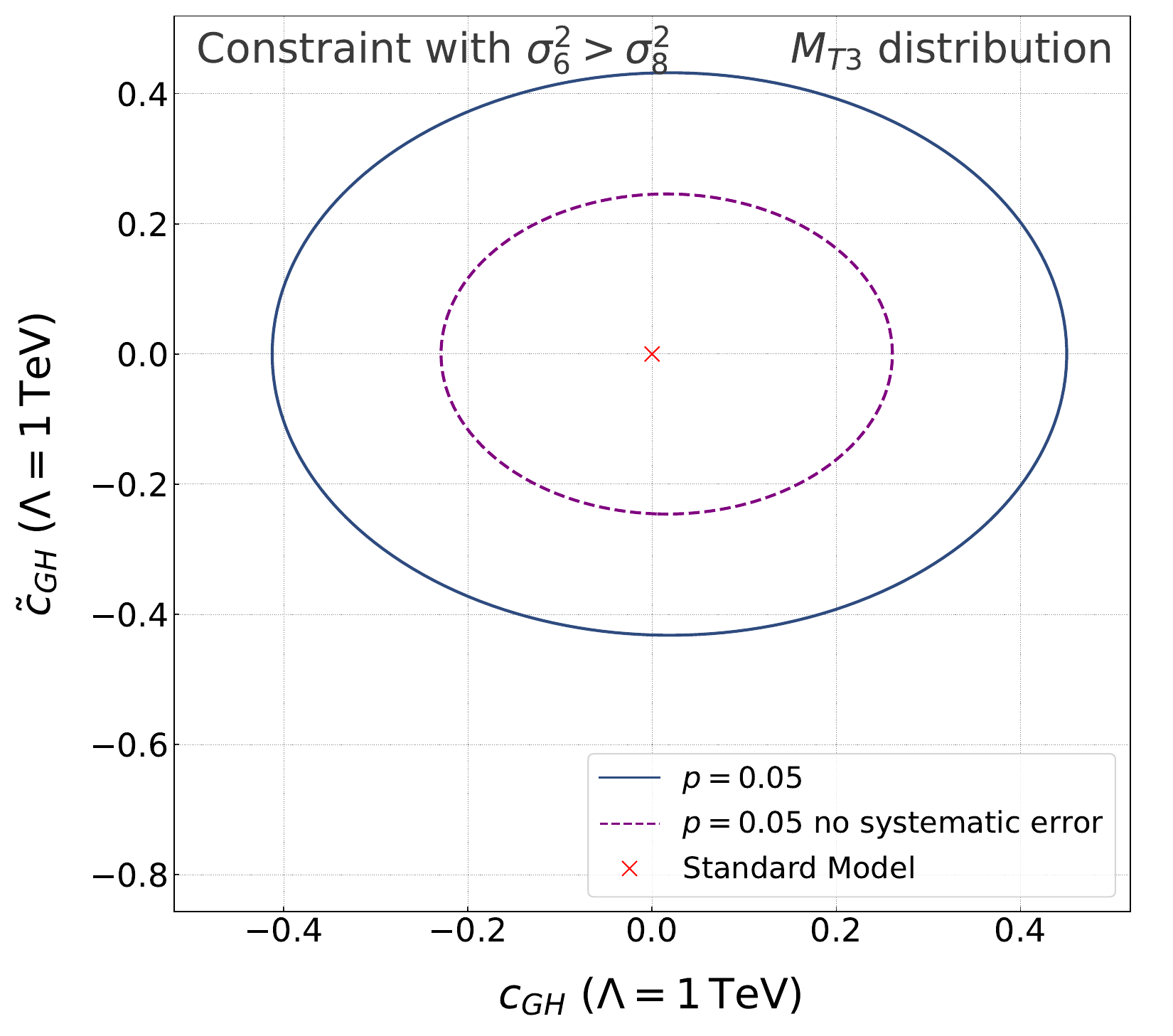}
  \caption{Sensitivity studies on the CP-even and CP-odd version of the 
  dimension-6 operator ($\phi\phi\, GG$) performed on projected HL-LHC data.
  EFT validity is ensured by only using bins for which dimension-6 squared contribution is larger than
  the dimension-8 squared contribution. On the left the $M_{e\mu}$ distribution is 
  used to make fits whereas on the right the $M_{T3}$ distribution is used.}
  \label{fig:contour_plot_comparisonM_{T3}}
\end{figure}

The CBB method automatically ensures the validity of the EFT
expansion. Once this is established, one can use the observable that
provides the best sensitivity. In this case, this is $M_{e\mu}$, and
not $M_{T3}$, despite the latter being better correlated to $M_{WW}$
(see the right panel of
figure~\ref{fig:contour_plot_comparisonM_{T3}}). This is because the
bins which are added in the EFT valid region for $M_{T3}$ are at lower
energies, where the sensitivity to higher dimensional operators is
less pronounced (see figure~\ref{fig:mt3_memu_comparison}). It is
because $M_{T3}$ is better correlated with $M_{WW}$ that these lower
energy regions receive much less BSM contributions than for
$M_{e\mu}$. It should also be noted that the $M_{T3}$ distribution
uses two fewer bins than the $M_{e\mu}$ distribution.
\newpage
\section{Conclusions}

In this letter, we have addressed the question of how to probe higher
dimensional operators in $WW$ production, while still ensuring that
the EFT expansion is valid. This is more challenging than in other
channels, because we do not have direct access to the invariant mass
of the $WW$ pair $M_{WW}$, which determines the relative size of different
orders in the EFT expansion. In particular, we have considered fully
leptonic $WW$ production via gluon fusion, and the distribution in the
dilepton invariant mass $M_{e\mu}$, which is used in all experimental
analyses.  As a case study, we probed the relative contributions of a
dimension-6 operator which added a $ggh$ contact interaction and a
dimension-8 operator which adds a $ggWW$ contact interaction, already
studied in~\cite{Gillies:2024mqp}. Both operators give amplitudes
which grow with energy, and can be constrained by studying the tails
of the $M_{e\mu}$ distribution. Note that a similar analysis can be
applied to higher-dimensional operators in the $q\bar q$ channel,
which will be subject of future work.

First, we have observed that, na\"ively imposing a cut on $M_{WW}$ at
the generator level to obtain information on the bins in $M_{e\mu}$
that can be safely used to obtain constraints on dimension-6
operators, does not eliminate the contribution of dimension-8
operators. In particular, there are bins in $M_{e\mu}$ where the
latter dominate, thus invalidating the EFT expansion. Its validity can
be restored only by manually selecting the bins where dimension-6
operators give the largest contribution. This unfortunately, restricts
the constraining power of such analyses.

This is ultimately due to the poor correlation between
$M_{e\mu}$ and $M_{WW}$. We have then considered three alternative
variables, $M_{T1}, M_{T2}$ and $M_{T3}$, and shown that $M_{T3}$ has
the best correlation with $M_{WW}$. Imposing a cut on $M_{T3}$ identifies the region in which the EFT is
valid in a way that can be implemented at the experimental level,
i.e.\ directly on the data. This in turn makes it possible to
investigate observables that have better sensitivity to higher
operators, while remaining the region of EFT validity.

We finally remark that adding $M_{WW} < \Lambda$ cuts only to the EFT
simulation could be interpreted as modifying the SMEFT expansion by a
form factor and could impact the model independence of EFT fits under
this procedure.
Instead, imposing a cut on $M_{T3}$ identifies the region in which the EFT is
valid in a way that can be implemented at the experimental level,
i.e.\ directly on the data. This in turn makes it possible to
investigate observables that have better sensitivity to higher
operators, while remaining the region of EFT validity.

\section*{Acknowledgements}
We thank Ennio Salvioni, Jonas Lindert, Tom Gent, Charlie Hogg, Xavier Pritchard, Marco
Sebastianutti, Lewis Mazzei, and Claudia Muni for useful discussions.
AB is supported by the UK STFC under the Consolidated Grant
ST/X000796/1, and thanks Royal Holloway, University of London for
hospitality while this work was completed. The work of AM is partially
supported by the National Science Foundation under Grant Number
PHY-2412701.
We acknowledge the use of computing resources made available by the Cambridge Service 
for Data Driven Discovery (CSD3), part of which is operated by the University of 
Cambridge Research Computing on behalf of the STFC DiRAC HPC Facility (www.dirac.ac.uk). 
The DiRAC component of CSD3 was funded by BEIS capital funding via STFC capital grants 
ST/P002307/1 and ST/R002452/1 and STFC operations grant ST/R00689X/1. DiRAC is part of 
the UK National e-Infrastructure.

\bibliography{main}

\begin{thebibliography}{10}

\bibitem{Brivio:2017vri}
Ilaria Brivio and Michael Trott.
\newblock {The Standard Model as an Effective Field Theory}.
\newblock {\em Phys. Rept.}, 793:1--98, 2019.

\bibitem{Grzadkowski:2010es}
B.~Grzadkowski, M.~Iskrzynski, M.~Misiak, and J.~Rosiek.
\newblock {Dimension-Six Terms in the Standard Model Lagrangian}.
\newblock {\em JHEP}, 10:085, 2010.

\bibitem{Brivio:2022pyi}
Ilaria Brivio et~al.
\newblock {Truncation, validity, uncertainties}.
\newblock 1 2022.

\bibitem{ElFaham:2024uop}
Hesham El~Faham, Giovanni Pelliccioli, and Eleni Vryonidou.
\newblock {Triple-gauge couplings in LHC diboson production: a SMEFT view from
  every angle}.
\newblock {\em JHEP}, 08:087, 2024.

\bibitem{Azatov:2019xxn}
A.~Azatov, D.~Barducci, and E.~Venturini.
\newblock {Precision diboson measurements at hadron colliders}.
\newblock {\em JHEP}, 04:075, 2019.

\bibitem{Brivio:2017btx}
Ilaria Brivio, Yun Jiang, and Michael Trott.
\newblock {The SMEFTsim package, theory and tools}.
\newblock {\em JHEP}, 12:070, 2017.

\bibitem{Azatov:2016sqh}
Aleksandr Azatov, Roberto Contino, Camila~S. Machado, and Francesco Riva.
\newblock {Helicity selection rules and noninterference for BSM amplitudes}.
\newblock {\em Phys. Rev. D}, 95(6):065014, 2017.

\bibitem{Degrande:2023iob}
C{\'e}line Degrande and Hao-Lin Li.
\newblock {Impact of dimension-8 SMEFT operators on diboson productions}.
\newblock {\em JHEP}, 06:149, 2023.

\bibitem{ElFaham:2025fow}
Hesham El~Faham, Giuseppe Ventura, and Eleni Vryonidou.
\newblock {Diboson production in the SMEFT at dimension-8}.
\newblock 11 2025.

\bibitem{Martin:2023tvi}
Adam Martin.
\newblock {A case study of SMEFT $ \mathcal{O}\left(1/{\Lambda}^4\right) $
  effects in diboson processes: pp {\textrightarrow}
  W$^{\pm}$({\ensuremath{\ell}}$^{\pm}${\ensuremath{\nu}}){\ensuremath{\gamma}}}.
\newblock {\em JHEP}, 05:223, 2024.

\bibitem{ATLAS:2025dhf}
Georges Aad et~al.
\newblock {Measurements of $W^+W^-$ production cross-sections in $pp$
  collisions at $\sqrt{s}=13$ TeV with the ATLAS detector}.
\newblock 5 2025.

\bibitem{CMS:2025nnv}
{Measurement of the photon-fusion production cross section of a pair of W
  bosons}.
\newblock 7 2025.

\bibitem{CMS:2020mxy}
Albert~M Sirunyan et~al.
\newblock {W$^+$W$^-$ boson pair production in proton-proton collisions at
  $\sqrt{s} =$ 13 TeV}.
\newblock {\em Phys. Rev. D}, 102(9):092001, 2020.

\bibitem{ATLAS:2019rob}
Morad Aaboud et~al.
\newblock {Measurement of fiducial and differential $W^+W^-$ production
  cross-sections at $\sqrt{s}=13$ TeV with the ATLAS detector}.
\newblock {\em Eur. Phys. J. C}, 79(10):884, 2019.

\bibitem{CMS:2018zzl}
Albert~M. Sirunyan et~al.
\newblock {Measurements of properties of the Higgs boson decaying to a W boson
  pair in pp collisions at $\sqrt{s}=$ 13 TeV}.
\newblock {\em Phys. Lett. B}, 791:96, 2019.

\bibitem{ATLAS:2017uhp}
Morad Aaboud et~al.
\newblock {Search for heavy resonances decaying into $WW$ in the $e\nu\mu\nu$
  final state in $pp$ collisions at $\sqrt{s}=13$ TeV with the ATLAS detector}.
\newblock {\em Eur. Phys. J. C}, 78(1):24, 2018.

\bibitem{ATLAS:2017bbg}
Morad Aaboud et~al.
\newblock {Measurement of the $W^+W^-$ production cross section in $pp$
  collisions at a centre-of-mass energy of $\sqrt{s}$ = 13 TeV with the ATLAS
  experiment}.
\newblock {\em Phys. Lett. B}, 773:354--374, 2017.

\bibitem{ATLAS:2017zuf}
Morad Aaboud et~al.
\newblock {Search for diboson resonances with boson-tagged jets in $pp$
  collisions at $\sqrt{s}=13$ TeV with the ATLAS detector}.
\newblock {\em Phys. Lett. B}, 777:91--113, 2018.

\bibitem{ATLAS:2017jag}
Morad Aaboud et~al.
\newblock {Search for $WW/WZ$ resonance production in $\ell \nu qq$ final
  states in $pp$ collisions at $\sqrt{s} =$ 13 TeV with the ATLAS detector}.
\newblock {\em JHEP}, 03:042, 2018.

\bibitem{Bellan:2021dcy}
Riccardo Bellan et~al.
\newblock {A sensitivity study of VBS and diboson WW to dimension-6 EFT
  operators at the LHC}.
\newblock {\em JHEP}, 05:039, 2022.

\bibitem{Banerjee:2024eyo}
Shankha Banerjee, Daniel Reichelt, and Michael Spannowsky.
\newblock {Electroweak corrections and EFT operators in W+W- production at the
  LHC}.
\newblock {\em Phys. Rev. D}, 110(11):115012, 2024.

\bibitem{Bellm:2016cks}
Johannes Bellm, Stefan Gieseke, Nicolas Greiner, Gudrun Heinrich, Simon
  Plätzer, Christian Reuschle, and Johann~Felix von Soden-Fraunhofen.
\newblock {Anomalous coupling, top-mass and parton-shower effects in ${W^+W^-}$
  production}.
\newblock {\em JHEP}, 05:106, 2016.

\bibitem{Falkowski:2016cxu}
Adam Falkowski, Martin Gonzalez-Alonso, Admir Greljo, David Marzocca, and Minho
  Son.
\newblock {Anomalous Triple Gauge Couplings in the Effective Field Theory
  Approach at the LHC}.
\newblock {\em JHEP}, 02:115, 2017.

\bibitem{Gillies:2024mqp}
Daniel Gillies, Andrea Banfi, Adam Martin, and Matthew~A. Lim.
\newblock {Dimension-8 operators in W+W- production via gluon fusion}.
\newblock {\em JHEP}, 06:111, 2025.

\bibitem{Arpino:2019fmo}
Luke Arpino, Andrea Banfi, Sebastian J\"ager, and Nikolas Kauer.
\newblock {BSM $WW$ production with a jet veto}.
\newblock {\em JHEP}, 08:076, 2019.

\bibitem{Campbell:1999ah}
John~M. Campbell and R.~Keith Ellis.
\newblock {An Update on vector boson pair production at hadron colliders}.
\newblock {\em Phys. Rev. D}, 60:113006, 1999.

\bibitem{Campbell:2011bn}
John~M. Campbell, R.~Keith Ellis, and Ciaran Williams.
\newblock {Vector Boson Pair Production at the LHC}.
\newblock {\em JHEP}, 07:018, 2011.

\bibitem{Campbell:2015qma}
John~M. Campbell, R.~Keith Ellis, and Walter~T. Giele.
\newblock {A Multi-Threaded Version of MCFM}.
\newblock {\em Eur. Phys. J. C}, 75(6):246, 2015.

\bibitem{Bertone:2017bme}
Valerio Bertone, Stefano Carrazza, Nathan~P. Hartland, and Juan Rojo.
\newblock {Illuminating the photon content of the proton within a global PDF
  analysis}.
\newblock {\em SciPost Phys.}, 5(1):008, 2018.

\bibitem{NNPDF:2021njg}
Richard~D. Ball et~al.
\newblock {The path to proton structure at 1{\%} accuracy}.
\newblock {\em Eur. Phys. J. C}, 82(5):428, 2022.

\end{thebibliography}

\end{document}